\documentclass[12pt,a4paper]{article}
\usepackage[utf8]{inputenc}
\usepackage{epsfig}
\usepackage{mathtools}
\usepackage{amsfonts}
\usepackage{amssymb}
\usepackage{amsmath}
\usepackage{cancel}
\usepackage{color}
\usepackage{slashed}
\usepackage{cite}
\usepackage{caption}
\usepackage{subcaption}
\usepackage{comment}
\usepackage{marginnote}
\usepackage{accents}

\usepackage{geometry} 
\usepackage{graphicx}

\usepackage{multirow}           
\usepackage{array}

\allowdisplaybreaks

\newcommand{\ep}{\varepsilon}
\newcommand{\OO}{\mathcal{O}}
\newcommand{\yy}{{1/y}}

\def\kite {{\rm kite}}

\begin{document}

\begin{center}
	
\vspace{2cm}		
		
{\Large\bf Massive kite diagrams with elliptics} \vspace{1cm}

{\large M.A. Bezuglov$^{1,2,3}$, A.I. Onishchenko$^{1,3,4}$, O.L. Veretin$^{5}$}\vspace{0.5cm}
		
{\it
$^1$Bogoliubov Laboratory of Theoretical Physics,  Joint
Institute for Nuclear Research, Dubna, Russia, \\
$^2$Moscow Institute of Physics and Technology (State University), Dolgoprudny, Russia, \\
$^3$Budker Institute of Nuclear Physics, Novosibirsk, Russia, \\
$^4$Skobeltsyn Institute of Nuclear Physics,  Moscow State University, Moscow, Russia, \\
$^5$Institute f\"ur Theoretische Physik, Universit\"at Regensburg, Regensburg, Germany}
\vspace{1cm}

\abstract{We present the results for two-loop massive kite master integrals with elliptics in terms of iterated integrals with algebraic kernels. The key ingredients are new integral representations for sunset subgraphs in $d=4-2\ep$ and $d=2-2\ep$ dimensions together with differential equations for considered kite master integrals in $A+B\ep$ form. The obtained results can be easily generalized to all orders in $\ep$-expansion and show that the class of functions defined as iterated integrals with algebraic kernels may be large enough for writing down results for a large class of massive Feynman diagrams.
}

\end{center}

\newpage

\section{Introduction}

Recently, we have seen a lot of progress in the evaluation of multiloop Feynman diagrams, in particular those with masses. The most impressive results in the realm of massive Feynman diagrams were obtained with the use of differential equation method \cite{diffeqn1,diffeqn2,diffeqn3,diffeqn4,diffeqn5}. In many cases the results can be written in terms of  multiple polylogarithms (MPLs) \cite{polylog1,polylog2,polylog3}. It turned out, that the whole possibility to have a polylogarithmic solution for the system of differential equations is tightly connected to the existence of its so called $\ep$-form  \cite{epform1,epform2}, see also \cite{epform-criterium} for a criterion of such reducibility.

Unfortunately, the class of functions given by multiple polylogarithms is not sufficient for calculating master integrals whose differential equations systems can not be reduced to $\ep$-form. Still, we already have a lot of progress in understanding simplest functions beyond multiple polylogarithms, the so-called elliptic polylogarithms \cite{Beilinson:1994,Wildeshaus,Levin:1997,Levin:2007,Enriquez:2010,Brown:2011,Bloch:2013tra,Adams:2014vja,Bloch:2014qca,Adams:2015gva,Adams:2015ydq,Adams:2016xah,Remiddi:2017har,Broedel:2017kkb,Broedel:2017siw,Broedel:2018iwv,Broedel:2018qkq,Broedel:2019hyg,Broedel:2019tlz,Bogner:2019lfa,Broedel:2019kmn,Walden:2020odh,Weinzierl:2020fyx}.
Further extensions can include cases with several elliptic curves \cite{Adams:2018bsn,Adams:2018kez}
or one can meet completely new functions, such as e.g in \cite{Bloch:2014qca,Primo:2017ipr,Bourjaily:2017bsb,Bourjaily:2018ycu,Bourjaily:2018yfy}.

In this paper we would like to show that the class of functions defined as iterated integrals with algebraic kernels can be large enough for a large enough class of physical problems, where one encounters evaluation of massive Feynman integrals. As a particular example we consider three different two-loop massive kite diagrams, which contain elliptics. Among these kite diagrams only one, the one with two massless lines, can be written in terms of elliptic polylogarithms, while the other two require introduction of more complicated structures,
which we can call hyperelliptic polylogarithms. These are the generalization of elliptic polylogarithms on algebraic curves with degrees higher than one. We would like to stress, that the proposed class of functions is different from iterated integrals with modular forms in kernels \cite{Adams:2015ydq,Adams:2017ejb,Ablinger:2017bjx}. To be able to write down results for kite diagrams in a chosen class of functions we used new integral representations for sunset subgraphs in $d=4-2\ep$ and $d=2-2\ep$ dimensions.
We consider the systems of differential equations for considered kite master integrals
and bring them to $A+B\ep$ form.
This form, being the natural extension of the already mentioned $\ep$-form, is so general, that
we expect it to be applicable to the most if not all differential systems, related to Feynman diagrams.
We give explicit results for two-loop sunsets up to $\OO(\ep^3)$ and kites up to $\OO (\ep^2)$ terms, but the obtained results can be easily generalized to any required order in $\ep$-expansion.

The paper is organized as follows. In the next section we present details on derivation of new integral representations for elliptic sunset subgraphs. Then, in section \ref{kites-sec} we show how the obtained integral representations for sunsets together with the reduction of kite's differential equations systems to $A+B\ep$ form can be used  to write down expressions for the latter in terms of iterated integrals with algebraic kernels. In the main body of the paper we used as example kites with sunset subgraphs in $d=2-2\ep$ dimension. In this case we have the most compact expressions for elliptic kite masters. Appendices explain our notation for iterated integrals and contain results for elliptic kite master integrals in the case of sunsets in $d=4-2\ep$ dimension. Finally, we have also prepared Mathematica notebook with all required details of calculation and results.

\section{Two-loop sunset diagrams}\label{sunset-sec}

In order to obtain results for massive kite master integrals in terms of iterated integrals with algebraic kernels, we used new integral representations for their elliptic sunset subgraphs.
This is the goal of this section to derive such representation for the two-loop massive sunset diagrams
with three equal masses, shown in Fig.~1.
We shall consider these diagrams in two diferent dimensions, $d=4-2\ep$ and $d=2-2\ep$, in the later subsections.

\begin{figure}[h]
	\centering
	\includegraphics[scale=0.6]{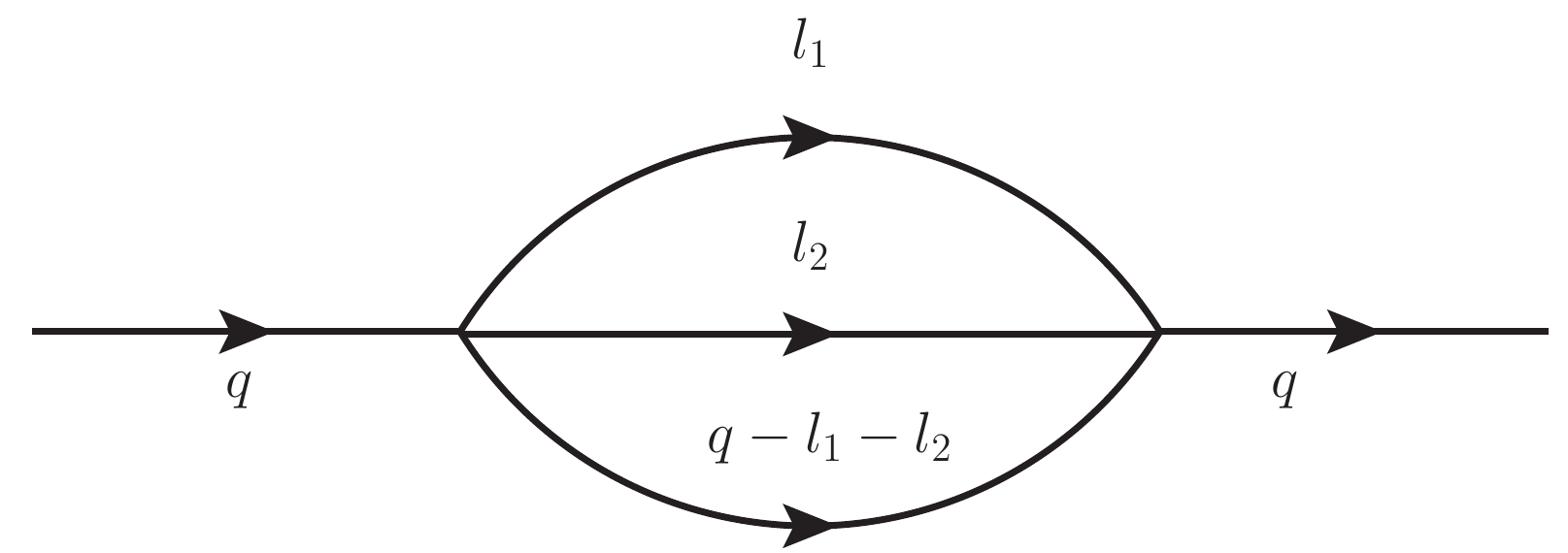}
	\caption{Sunset diagram.}
	\label{fig:sunset}
\end{figure}

It is well known that we need three master integrals to close up the system of differential
equations for the sunset with equal masses.
Namely, let us introduce the following notation for our set of three
master integrals in $d$ dimensions\footnote{See Fig. \ref{fig:sunset} for distribution of momenta.} $(n=0,1,2)$
\begin{equation}
  \label{eq:S_def}
S_{n,1,1}(q^{2})=\int\frac{dl_{1}dl_{2}}{(i\pi^{d/2})^{2}}\frac{1}{(1-l_{1}^{2})^{n}(1-l_{2}^{2})(1-l_{3}^{2})}\,,
\end{equation}
where $l_{3}=q-l_{1}-l_{2},\;q^{2}=s$ and we set all masses to the unity $m=1$.
Integral $S_{0,1,1}(q^2)$ is just
a product of one-loop bubble integrals while the other two are non-trivial.

First we use Feynman parameter trick for
the last two propagators $1/(1-l_2^2)$ and $1/(1-l_3^2)$,
as was introduced in \cite{KKOVelliptic1,KKOVelliptic2,LinearReducibledEllipticFeynmanIntegrals},
and rewrite Eq.~(\ref{eq:S_def}) as
\begin{equation}
  S_{n,1,1}(q^2)=\int_{0}^{1}dt\int\frac{dl_{1}dl_{2}}{(i\pi^{d/2})^{2}}
     \frac{1}{(1-l_{1}^{2})^{n}(1-t\bar{t}\big(l_{1}-q)^{2}-l_{2}^{2}\big)^{2}} \,,
\end{equation}
where $\bar{t}=1-t$. We can now integrate over loop momentum $l_2$ and obtain
\begin{equation}
  \label{eq:S_y}
  S_{n,1,1}(q^2)=\Gamma(2-d/2)\int_{0}^{1}dt\,(t\bar{t})^{d/2-2}\int\frac{dl_{1}}{i\pi^{d/2}}
  \frac{1}{(1-(l_{1}+q)^{2})^{n}(y^{2}-l_{1}^{2})^{2-d/2}} \,.
\end{equation}
Here we introduced the short hand notation $y^{2}=1/(t\bar{t})$, which
plays a r\^ole of a new mass in a one-loop subgraph.
The expressions for one-loop subgraphs in $l_1$ will be denoted as
\begin{equation}
  \label{eq:In}
I_{n}=\int\frac{dl_{1}}{i\pi^{d/2}}\frac{1}{(1-(l_{1}+q)^{2})^{n}(y^{2}-l_{1}^{2})^{2-d/2}} \,.
\end{equation}

The idea now is to apply differential equations method \cite{diffeqn1,diffeqn2,diffeqn3,diffeqn4,diffeqn5}
for the one-loop diagram in representation (\ref{eq:In}).

In the next two subsections we consider $\ep$-expansions of the above formulae near $d$ equals 4 and 2
respectively.
In order to distinguish the results we shall supply all relevant quantities with the
superscript and write, e.g., $S^{(4)}_{n,1,1}$ instead of $S_{n,1,1}$ and so on.

\subsection{Case $d=4-2\ep$}

The expressions for one-loop integral (\ref{eq:In})
can be conveniently obtained using differential equation method.
First, using integration by parts (IBP) identities \cite{IBP1,IBP2} we write  two $3\times3$ systems of differential equations the over $y^2$ and $s$ 
in the basis $I^{(4)}=(I_1^{(4)},I_2^{(4)},I_3^{(4)})^\top$.
Further, by choosing a suitable basis $\widetilde{I}^{(4)}$ these systems
can be rewritten in $\ep$-form\footnote{The derivation of IBP identities and subsequent reduction of system of differential equations to $\ep$-form were performed with the use of  LiteRed \cite{LiteRed1,LiteRed2}  and Libra \cite{Libra} packages.} \cite{epform1,epform2}
\begin{align}
\frac{d\widetilde{I}^{(4)}}{dz} & =\varepsilon M_{z}^{(4)}\widetilde{I}^{(4)}, \label{I4-1}\\
\frac{d\widetilde{I}^{(4)}}{dy} & =\varepsilon M_{y}^{(4)}\widetilde{I}^{(4)}, \label{I4-2}
\end{align}
where we introduced new variable $z$, being the root of equation $s=(y-z)(y-1/z).$
Elements of new basis are linearly expressed through the old basis elements by means of 
transformation matrix $T_{\rm sunset}^{(4)}$, such that $I^{(4)}=T_{\rm sunset}^{(4)}\widetilde{I}^{(4)}$.

The matrices $M_{z}^{(4)}$ and $M_{y}^{(4)}$ are then given by 
\begin{align}
M_{z}^{(4)} & =\frac{M_{z,1}^{(4)}}{z}+\frac{M_{z,2}^{(4)}}{z-1}+\frac{M_{z,3}^{(4)}}{z+1}+\frac{M_{z,4}^{(4)}}{z-1/y}+\frac{M_{z,5}^{(4)}}{z-y}\,,\\
M_{y}^{(4)} & =\frac{M_{y,1}^{(4)}}{y}+\frac{M_{y,2}^{(4)}}{y-1/z}+\frac{M_{y,3}^{(4)}}{y-z} \,,
\end{align}
where
\begin{align}
M_{z,1}^{(4)} & =\left(\begin{array}{ccc}
0 & 0 & 0\\
-\frac{81}{5} & -2 & -2\\
\frac{243}{5} & 6 & 5
\end{array}\right),\,M_{z,2}^{(4)}=\left(\begin{array}{ccc}
0 & 0 & 0\\
0 & 2 & 2\\
0 & -6 & -6
\end{array}\right),\,M_{z,3}^{(4)}=\left(\begin{array}{ccc}
0 & 0 & 0\\
0 & 2 & 2\\
0 & -6 & -6
\end{array}\right),\,\\
M_{z,4}^{(4)} & =\left(\begin{array}{ccc}
0 & 0 & 0\\
0 & -3 & -2\\
0 & 6 & 4
\end{array}\right),\,M_{z,5}^{(4)}=\left(\begin{array}{ccc}
0 & 0 & 0\\
0 & 0 & 0\\
0 & 0 & 1
\end{array}\right),\\
M_{y,1}^{(4)} & =\left(\begin{array}{ccc}
-4 & 0 & 0\\
\frac{243}{5} & 2 & 2\\
-\frac{243}{5} & -6 & -6
\end{array}\right),\,M_{y,2}^{(4)}=\left(\begin{array}{ccc}
0 & 0 & 0\\
0 & -3 & -2\\
0 & 6 & 4
\end{array}\right),\,M_{y,3}^{(4)}=\left(\begin{array}{ccc}
0 & 0 & 0\\
0 & 0 & 0\\
0 & 0 & 1
\end{array}\right).
\end{align}
It should be noted here, that if we have had written differential equation w.r.t. $s$ rather than w.r.t.
new variable $z$, the above formulas would be much more complicated and include square roots.
This is the reason why we prefer to work with $z$ rather that with $s$.
Indeed the value of $z$ correspond to two possible solutions of the equation $s=(y-z)(y-1/z)$
and we can choose one of them, namely
\begin{align}
z(s,y) = \frac{1-s+y^2+\sqrt{(s-1-y^2)^2-4y^2}}{2y} \,. \label{zexpression}
\end{align}

The explicit expression for transformation matrix $T_{\rm sunset}^{(4)}$ is quite cumbersome and  may be found in Mathematica notebook accompanying this article. Next, we solve this
system of equations with boundary conditions at 
$s=0$.  In terms of variable $z$ these boundary conditions can be imposed either at  $z=y\, ~\textrm{or}~ z=1/y$. With our choice of $z$ in Eq.\eqref{zexpression} we chose them at $z=y$.

To find these boundary conditions we write and solve differential
equations for $(n=0,1,2)$
\begin{equation}
I_{b,n}^{(4)}=\int\frac{dl_{1}}{i\pi^{d/2}}\frac{1}{(1-l_{1}^2)^{n}(y^{2}-l_{1}^{2})^{\varepsilon}}
\end{equation}
integrals. 
This system could be also reduced to $\ep$-form and we have $((I_{b,1}^{(4)},I_{b,2}^{(4)},I_{b,3}^{(4)})^{\top}=T_{{\rm sunset},b}^{(4)}\widetilde{I}_b^{(4)})$:
\begin{equation}
\frac{d\widetilde{I}_b}{dy}=\ep M_b^{(4)}\widetilde{I}_b\,,
\end{equation}
where $M_b^{(4)}=\frac{M_{b,1}^{(4)}}{y}+\frac{M_{b,2}^{(4)}}{y-1}+\frac{M_{b,3}^{(4)}}{y+1}$ and 
\begin{equation}
M_{b,1}^{(4)}=\left(\begin{array}{ccc}
-4 & 0 & 0\\
0 & 0 & 0\\
0 & 0 & 0
\end{array}\right),\, M_{b,2}^{(4)}=\left(\begin{array}{ccc}
0 & 0 & 0\\
3 & -1 & 0\\
0 & 0 & -3
\end{array}\right),\, M_{b,3}^{(4)}=\left(\begin{array}{ccc}
0 & 0 & 0\\
3 & -1 & 0\\
0 & 0 & -3
\end{array}\right).
\end{equation}
The transformation matrix $T_{{\rm sunset},b}$ to this form is given by 
\begin{equation}
T_{{\rm sunset},b}=\left(\begin{array}{ccc}
y^{4} & 0 & 0\\
-2y^{2} & \frac{2(2\varepsilon-1)}{3\varepsilon} & (y-1)^{2}(y+1)^{2}\\
\frac{2y^{2}(2\varepsilon-1)}{(y-1)(y+1)} & \frac{2(2\varepsilon-1)\left(\varepsilon y^{2}-y^{2}-2\varepsilon+1\right)}{3(y-1)(y+1)\varepsilon} & -(y-1)(y+1)(3\varepsilon-2)
\end{array}\right).
\end{equation}
The obtained system is then easily solved in terms of multiple polylogarithms using the following boundary conditions at $y=0$:
\begin{align}
\int\frac{dl_{1}}{i\pi^{\nu}}\frac{1}{(1-l_{1}^2)^{n}(-l_{1}^{2})^{\varepsilon}} & =\frac{\Gamma(n+2\varepsilon-2)\Gamma(2-2\varepsilon)}{\Gamma(n)\Gamma(2-\varepsilon)},\:n>0\\
\int\frac{dl_{1}}{i\pi^{\nu}}\frac{1}{(y^{2}-l_{1}^{2})^{\varepsilon}} & =\frac{\Gamma(2\varepsilon-2)}{\Gamma(\varepsilon)}y^{4-4\varepsilon}
\end{align}
This way we get\footnote{See for notation Appendix \ref{notation}.}
\begin{align}
e^{\ep\gamma_E}I_{b,1}^{(4)} &= \frac{y^4}{4}+\ep\left(\frac{3 y^4}{4}-y^4 \log (y)\right) + \OO (\ep^2) , \\
e^{\ep\gamma_E}I_{b,2}^{(4)} &= -\frac{1}{2\ep}-\frac{1}{2} \left(y^2+1\right) \nonumber \\
 & + \ep\left(2 G(-1,0;y)+2 G(1,0;y)+2 y^2 \log{y} - \frac{3}{2}y^2 - \frac{7}{24}\pi^2 -\frac{1}{2} \right) + \OO (\ep^2) , \\
e^{\ep\gamma_E}I_{b,3}^{(4)} &= \frac{1}{2\ep}-\frac{1}{2} \nonumber \\ & +\ep\left(-2 G(-1,0;y)-2 G(1,0;y)+\frac{2 y^2 \log{y}}{(y-1) (y+1)}+ \frac{7}{24} \pi
^2- \frac{1}{2}\right)  + \OO (\ep^2) \,,
\end{align}
where $\gamma_E=0.5772156649\dots$ is the Euler--Mascheroni constant.

Having determined boundary conditions for $I_n^{(4)}$ integrals at $z=y$ the solution for differential system in Eq. \eqref{I4-1} is also straightforwardly written in terms of multiple polylogarithms. Using the latter one immediately gets the desired expressions for original two-loop sunset master integrals. Note, that $S_{0,1,1}^{(4)}(q^2)$ master integral is the product of two one-loop tadpoles and it is more convenient to find its expression this way without performing the last integration over Feynman parameter $t$. Moreover, its integral representation above is divergent and thus ill defined. Also, careful inspection\footnote{One can also simply check its numerical convergence.} shows that the integral representation for $S_{1,1,1}^{(4)}(q^2)$ master integral is also ill defined. To remedy this problem we take another master integral $S_{3,1,1}^{(4)}(q^2)$ as part of our basis of master integrals and calculate its expression from IBP relations for $I_3^{(4)}$ one-loop subgraph. Altogether we get:
\begin{align}
e^{2\ep\gamma_E} S_{2,1,1}^{(4)} =& \frac{1}{2\ep^2} + \frac{1}{2\ep} + \frac{\pi^2}{6} - \frac{3}{2} + \int_2^{\infty}\frac{4\, dy}{y^2\sqrt{y^2-4}}\widetilde{S}_2(y,z) + \OO (\ep), \\
e^{2\ep\gamma_E} S_{3,1,1}^{(4)} =& \frac{1}{2\ep} + \int_2^{\infty}\frac{4\,dy}{y^2\sqrt{y^2-4}}\widetilde{S}_3 (y,z) + \OO (\ep),
\end{align} 
where ($G$-functions stand for usual multiple polylogarithms, see also Appendix A)
\begin{align}
\widetilde{S}_2(y,z) =& G(y,0;z) - G(1/y,0;z) + \frac{\log^2 y}{2} - \frac{\log^2 z}{2} +\left(1+\frac{y}{z-y}+\frac{1}{yz-1}\right)\log z \nonumber \\
&  + \left(
\frac{1}{yz-1} - \frac{y}{z-y} + \log z - G(y;z) - G(1/y;z)
\right)\log y\, , \\
\widetilde{S}_3(y,z) =& \frac{(2z+y^2z-y(1+z^2))}{2(y-z)(yz-1)}\log y + \frac{y (1+z^4-yz(1+z^2))}{2(y-z)(z^2-1)(yz-1)}\log z\, .
\end{align}
Alternatively, using notation from Appendix \ref{notation} the same expressions read
\begin{align}
\label{Sunset4d}
e^{2\ep\gamma_E} S_{2,1,1}^{(4)} =& \frac{1}{2 \varepsilon^2}+\frac{1}{2 \varepsilon }+\frac{\pi ^2}{6}-\frac{3}{2} \nonumber  \\
& + J\left(\Theta _\yy,\omega _0^y;s\right)-J\left(\Omega _0,\omega _\yy^z,\omega _0^y;s\right)+J\left(\Theta _\yy,\omega
_0^z;s\right) -J\left(\Omega _0,\omega _\yy^z,\omega _0^z;s\right)
\nonumber  \\ &
-J\left(\Theta _y,\omega _0^y;s\right)+J\left(\Omega
_0,\omega _0^y,\omega _0^y;s\right)+J\left(\Theta _y,\omega _0^z;s\right)  +J\left(\Omega _0,\omega _0^z,\omega
_0^y;s\right) \nonumber 
\\ &
-J\left(\Omega _0,\omega _y^z,\omega _0^y;s\right)  +J\left(\Omega _0,\omega _y^z,\omega
_0^z;s\right)
+J\left(\Omega _0,\omega _0^z;s\right)-J\left(\Omega _0,\omega _0^z,\omega _0^z;s\right)\, , \\
e^{2\ep\gamma_E} S_{3,1,1}^{(4)} =& \frac{1}{2 \varepsilon } +
\frac{1}{2} \left(1-J\left(\Theta _1,\omega _0^y;s\right)+J\left(\Theta _2,\omega _0^y;s\right)+J\left(\Theta _1,\omega
_0^z;s\right)\right. \nonumber \\ &\left.
+J\left(\Theta _2,\omega _0^z;s\right)-J\left(\Theta _3,\omega _0^z;s\right)+J\left(\Theta _4,\omega
_0^z;s\right)-J\left(\Omega _0,\omega _0^z;s\right)\right)\, .
\end{align}
We have checked that the obtained integral representations for considered sunset master integrals do correctly reproduce known numerical expressions\footnote{For numerical cross-checks we used sector decomposition method \cite{Binoth:2000ps,Binoth:2003ak,Binoth:2004jv,Heinrich:2008si,Bogner:2007cr,Bogner:2008ry,Kaneko:2009qx} as implemented in \cite{Fiesta4}. } both below and above threshold. In addition to the expressions above the Mathematica notebook coming together with this paper contains expressions for further corrections in $\ep$ up to $\OO (\ep^3)$.

Note, that the most important feature of this representation is that its dependence on kinematic variable $s$ is only through the upper limit of penultimate integration and $\Theta$ 1-forms participating in the final integration over $y$. This property will allow us to further substitute obtained sunset expressions into differential equations over $s$ for higher master integrals having sunsets in their subgraphs. It is precisely what we want do for kite diagrams in the next section.

\subsection{Case $d=2-2\ep$}

Following the same steps as in the previous subsection we can obtain expressions for sunset master integrals in $d=2-2\ep$ dimension. The latter are given by
\begin{align}
e^{2\ep\gamma_E} S_{1,1,1}^{(2)} =& 8\int_2^{\infty}\frac{dy}{y^2\sqrt{y^2-4}}\frac{yz\log z}{(z^2-1)} + \OO(\ep)\, ,\\
e^{2\ep\gamma_E} S_{2,1,1}^{(2)} =& 4\int_2^{\infty}\frac{dy}{y^2\sqrt{y^2-4}}\left[
\frac{z(yz^2-2z+y)}{(z^2-1)^2} + \frac{2z^2(z^2-2yz+1)}{(z^2-1)^3}
\right] + \OO(\ep)\, .
\end{align} 
Using the notation from Appendix \ref{notation} these expressions can be written as
\begin{align}
e^{2\ep\gamma_E} S_{1,1,1}^{(2)} =& 2 J\left(\Theta _6,\omega _0^z; s\right) + \OO(\ep)\, ,\\
e^{2\ep\gamma_E} S_{2,1,1}^{(2)} =& -J\left(\Theta _5;s\right)+J\left(\Theta _6;s\right)-J\left(\Theta _8;s\right) \nonumber \\ & +\frac{1}{2} J\left(\Theta _5,\omega
_0^z;s\right)+\frac{3}{2} J\left(\Theta _8,\omega _0^z;s\right)-J\left(\Theta _9,\omega _0^z;s\right) + \OO(\ep)\, .
\end{align}
Further terms of expansion in $\ep$ for considered sunset master integrals may be found in accompanying Mathematica notebook.

The sunset master integrals in $d=4-2\ep$ and $d=2-2\ep$ dimensions are not independent from each other and can be related with each other by means of dimensional recurrence relations\cite{dimrecurrence}. This means that both of these representations can be used as basis elements for broader families of master integrals such as kites.

\section{Kite diagrams}\label{kites-sec}

We have seen in previous section that the integral representations for sunsets are simpler in $d=2-2\ep$ dimension
as compared to $d=4-2\ep$.
Therefore, in this section we shall consider results for kite diagrams in $d=4-2\ep$
using $(2-2\ep)$-dimensional sunsets.
The results for kite master integrals with $(4-2\ep)$-dimensional sunsets up to ${\cal O}(\ep)$
are given in Appendix B and higher terms in $\ep$-expansion
can be found in accompanying Mathematica notebook.

\subsection{Kite with two massless lines}

\begin{figure}[h]
	\centering
	\includegraphics[scale=0.6]{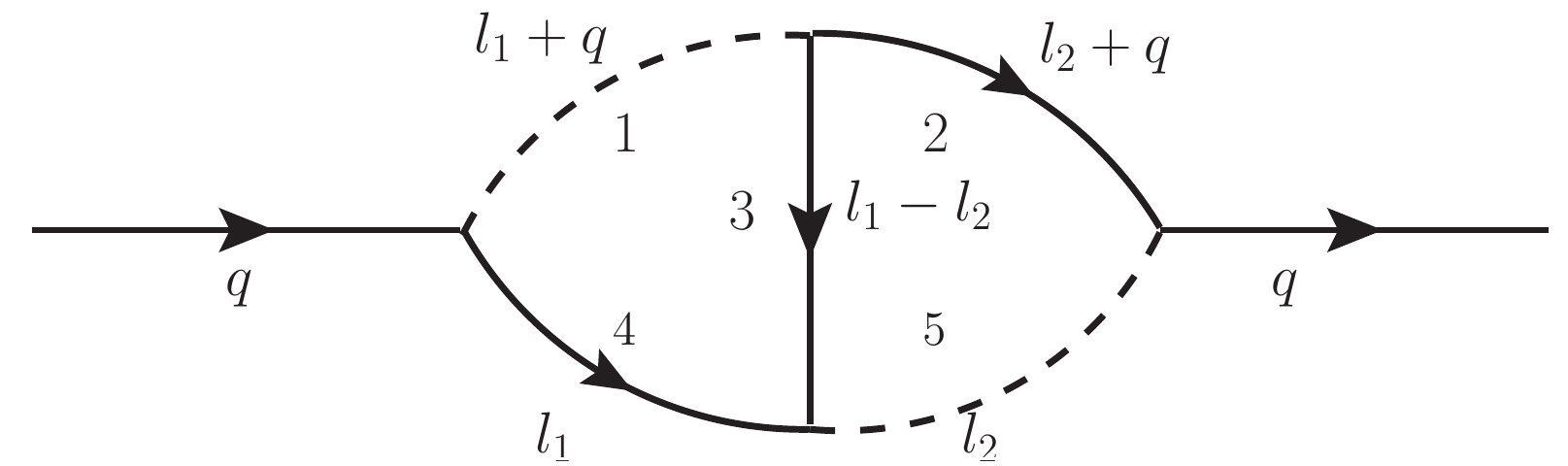}
	\caption{Kite diagram with two massless lines.}
	\label{fig:kite1}
\end{figure}

The simplest case of kite integrals containing elliptics is given by kite diagrams with two massless lines depicted in Fig. \ref{fig:kite1}. The master integrals in this family are defined as ($d=4-2\ep$):
\begin{equation}
j_1^{\kite}(a_1,\ldots , a_5) = \int\frac{d^d l_1 d^d l_2}{(i\pi^{d/2})^2}
\frac{1}{D_1^{a_1}D_2^{a_2}D_3^{a_3}D_4^{a_4}D_5^{a_5}},
\end{equation}
where
\begin{align}
D_1 &= -(l_1+q)^2,& D_2 &= 1-(l_2+q)^2,& D_3 &= 1-(l_1-l_2)^2,& D_4 &= 1-l_4^2,& D_5 &= -l_2^2\, . \nonumber \\
\end{align}
The vector of Laporta master integrals obtained as a result of IBP reduction  \cite{IBP1,IBP2} together with dimension reduction \cite{dimrecurrence} for sunset masters can be chosen in the following form: 
\begin{align}
I_{\rm Laporta} = \{
&j_1^{\kite}(0,0,1,1,0),& &j_2^{\kite}(0,1,0,1,1),& &S_{1,1,1}^{(2)},& &S_{2,1,1}^{(2)}, \nonumber \\ 
& j_1^{\kite}(1,0,1,0,1),& &j_1^{\kite}(1,0,1,0,2),& &j_1^{\kite}(1,1,0,1,1),& &j_1^{\kite}(1,1,1,1,1) 
\}^{\top}
\end{align}
 
\noindent
To evaluate these master integrals we consider their system of differential equations with respect to $q^2=s$. Using balance transformations of \cite{epform2} the latter can be reduced to the following $A+B\ep$ form\footnote{We used their implementation in  Libra package \cite{Libra}.}:
\begin{equation}
\frac{d\tilde{I}_{\rm canonical}}{ds} = M\tilde{I}_{\rm canonical}\, ,
\end{equation}
where 
\begin{equation}
M = \frac{M_1}{s-9} + \frac{M_2}{s-1} + \frac{M_3}{s}\, ,
\end{equation}
and 
\begin{align}
M_1 &= \left(
\begin{array}{cccccccc}
0 & 0 & 0 & 0 & 0 & 0 & 0 & 0 \\
0 & 0 & 0 & 0 & 0 & 0 & 0 & 0 \\
0 & 0 & 1 & 0 & 0 & 0 & 0 & 0 \\
0 & 0 & \frac{1}{12} (3 \varepsilon +1) & -2 \varepsilon  & 0 & 0 & 0 & 0 \\
0 & 0 & 0 & 0 & 0 & 0 & 0 & 0 \\
0 & 0 & 0 & 0 & 0 & 0 & 0 & 0 \\
0 & 0 & 0 & 0 & 0 & 0 & 0 & 0 \\
0 & 0 & 0 & 0 & 0 & 0 & 0 & 0 \\
\end{array}
\right) , \\
M_2 &= \left(
\begin{array}{cccccccc}
0 & 0 & 0 & 0 & 0 & 0 & 0 & 0 \\
0 & -2 \varepsilon  & 0 & 0 & 0 & 0 & 0 & 0 \\
0 & 0 & 0 & 0 & 0 & 0 & 0 & 0 \\
-2 \varepsilon  & 0 & \frac{1}{4} (3 \varepsilon +1) & -2 \varepsilon -1 & 0 & 0 & 0 & 0 \\
0 & 0 & 0 & 0 & 0 & 0 & 0 & 0 \\
0 & 0 & 0 & 0 & 0 & -4 \varepsilon  & 0 & 0 \\
0 & 0 & 0 & 0 & 0 & 0 & -4 \varepsilon  & 0 \\
\frac{1}{2} (-2 \varepsilon -1) & 0 & \frac{\varepsilon }{6} & 0 & \frac{4}{9} (2 \varepsilon +1) & 0 & 0 & -2 \varepsilon  \\
\end{array}
\right) , \\
M_3 &= \left(
\begin{array}{cccccccc}
0 & 0 & 0 & 0 & 0 & 0 & 0 & 0 \\
\frac{\varepsilon }{6} & \varepsilon  & 0 & 0 & 0 & 0 & 0 & 0 \\
0 & 0 & -2 \varepsilon -1 & 3 (2 \varepsilon +1) & 0 & 0 & 0 & 0 \\
0 & 0 & -\frac{1}{3} (3 \varepsilon +1) & 3 \varepsilon +1 & 0 & 0 & 0 & 0 \\
0 & 0 & 0 & 0 & -\frac{\varepsilon }{2} & -\frac{3 \varepsilon }{2} & 0 & 0 \\
0 & 0 & 0 & 0 & \frac{\varepsilon }{2} & \frac{3 \varepsilon }{2} & 0 & 0 \\
0 & \frac{\varepsilon }{3} & 0 & 0 & 0 & 0 & 2 \varepsilon  & 0 \\
\frac{1}{4} (2 \varepsilon +1) & 3 (2 \varepsilon +1) & 0 & 0 & 0 & 0 & 9 (2 \varepsilon +1) & \varepsilon  \\
\end{array}
\right) . 
\end{align}
The transformation matrix $T$ ($I_{\rm Laporta}=T\tilde{I}_{\rm canonical}$) to the canonical basis $\tilde{I}_{\rm canonical}$ can be found in accompanying Mathematica notebook. Having obtained the differential system in this form it is easy to see, that the solution for all master integrals except sunsets can be obtained recursively in the regularization parameter $\ep$ similarly to what one typically does for differential systems reduced to $\ep$-form. The solutions for sunsets themselves where already presented in the previous section. This way, accounting to boundary conditions at $s=0$ we get\footnote{At $s=0$ we need to calculate 2-loop tadpoles, which is a trivial problem at present.}:
\begin{multline}
e^{2\ep\gamma_E}j_1^{\kite}(1,1,1,1,1) =\frac{1}{6s}\Bigg(\pi ^2 G(1;s)+12 G(0,1,1;s)-6 G(1,0,1;s)+J\left(\Omega _1,\omega _{-1}^{\text{sz}},\omega _0^z;s\right) \\ +J\left(\Omega _1,\omega _1^{\text{sz}},\omega
	_0^z;s\right)-8 J\left(\Omega _1,\omega _{1,-1}^s,\omega _0^z;s\right)-8 J\left(\Omega _1,\omega _{1,1}^s,\omega _0^z;s\right)\Bigg) + \OO(\ep).
\end{multline}
The results for other master integrals together with ${\cal O}(\ep^1)$ term for kite $j_1^{\kite}(1,1,1,1,1)$ itself may be found in accompanying Mathematica file. Note, that in principle with the presented procedure we can have as many terms in $\ep$ expansion of considered master integrals as required. We would like also to point out that the result we got here is more compact then those previously obtained in \cite{Adams:2016xah,Remiddi:2016gno,Broedel:2019hyg}

\subsection{Kite with one massless line}

\begin{figure}[h]
	\centering
	\includegraphics[scale=0.6]{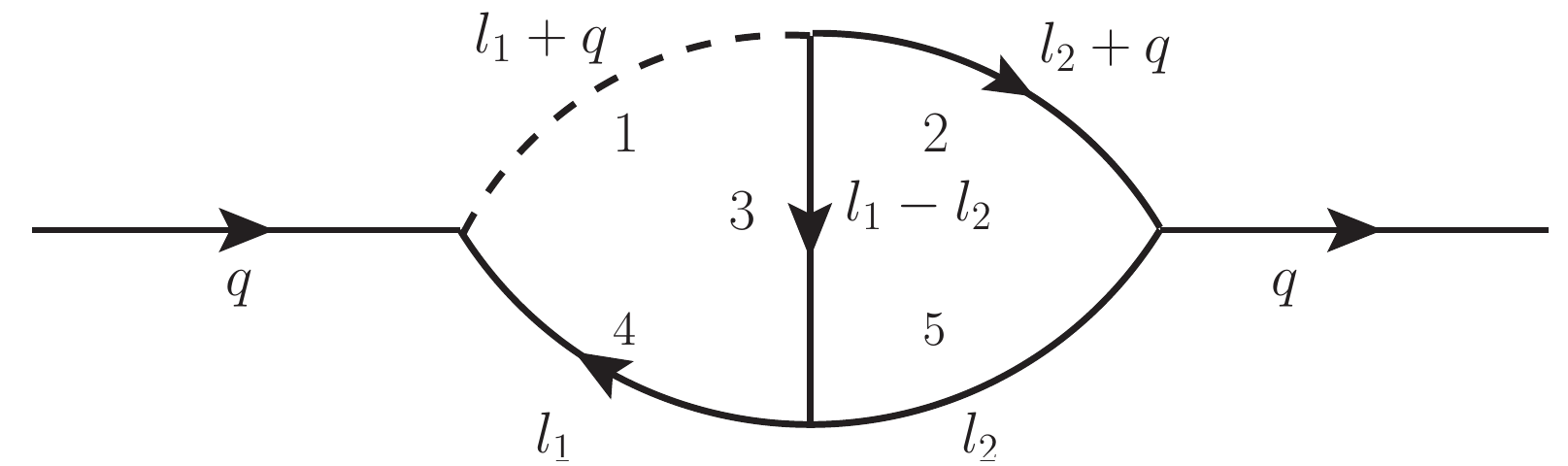}
	\caption{Kite diagram with one massless line.}
	\label{fig:kite2}
\end{figure}

Similar to the previous case we may also consider the kite diagram with only one massless line, see Fig. \ref{fig:kite2}.  The master integrals in this family are defined as ($d=4-2\ep$):
\begin{equation}
j_2^{\kite}(a_1,\ldots , a_5) = \int\frac{d^d l_1 d^d l_2}{(i\pi^{d/2})^2}
\frac{1}{D_1^{a_1}D_2^{a_2}D_3^{a_3}D_4^{a_4}D_5^{a_5}},
\end{equation}
where
\begin{align}
D_1 &= -(l_1+q)^2,& D_2 &= 1-(l_2+q)^2,& D_3 &= 1-(l_1-l_2)^2,& D_4 &= 1-l_4^2,& D_5 &= 1-l_2^2\, . \nonumber \\
\end{align}
The vector of Laporta master integrals in this case was chosen as: 
\begin{align}
I_{\rm Laporta} = \{
&j_2^{\kite}(0,0,0,1,1),& &j_2^{\kite}(0,0,1,1,1),& &j_2^{\kite}(0,1,0,1,1),& &S_{1,1,1}^{(2)},\nonumber \\  &S_{2,1,1}^{(2)},& &j_2^{\kite}(1,0,0,1,1),& &j_2^{\kite}(1,0,1,0,1),& &j_2^{\kite}(1,0,1,0,2),\nonumber \\
&j_2^{\kite}(0,1,1,1,1),& &j_2^{\kite}(1,0,1,1,1),& &j_2^{\kite}(1,1,0,1,1),& &j_2^{\kite}(1,1,1,1,1)
\}^{\top}
\end{align}

\noindent
The system of differential equations with respect to $q^2=s$ may be again reduced to $A+B\ep$ canonical form and we have ($I_{\rm Laporta}=T\tilde{I}_{\rm canonical}$):
\begin{equation}
\frac{d\tilde{I}_{\rm canonical}}{ds} = M\tilde{I}_{\rm canonical}\, ,
\end{equation}
where ($z_1 = \sqrt{\frac{4-s}{s}}$):
\begin{equation}
M = \frac{M_1}{s-9} + \frac{M_2}{s-4} + \frac{M_3}{s-1} + \frac{M_4}{s} + \frac{z_1 M_5}{s-9} + \frac{z_1 M_6}{s-4} + \frac{z_1 M_7}{s-1}
\end{equation}
and the particular expressions for coefficient matrices $M_i$ together with transformation matrix to canonical basis $T$ can be found in accompanying Mathematica notebook. The solution of differential system in the canonical basis goes similar to the case of kite diagram with two massless lines and for masters with elliptics different from sunset diagrams we get\footnote{See notation in Appendix \ref{notation}.}
\begin{multline}
e^{2\ep\gamma_E}j_2^{\kite}(1,1,1,1,1) = \frac{1}{s}\Big[-\frac{5}{4} J\left(\Omega _0,\omega _1^{z_1},\omega _{4,-1}^{z_1};s\right)+\frac{5}{4} J\left(\Omega _0,\omega _1^{z_1},\omega _{4,1}^{z_1};s\right)+\frac{5}{4}
J\left(\Omega _0,\omega _1^{z_1},\zeta _{4,-1}^{z_1};s\right) \\ +\frac{5}{4} J\left(\Omega _0,\omega _1^{z_1},\zeta _{4,1}^{z_1};s\right)+\frac{5}{4} J\left(\Omega
_0,\omega _4^{z_1},\omega _{4,-1}^{z_1};s\right)-\frac{5}{4} J\left(\Omega _0,\omega _4^{z_1},\omega _{4,1}^{z_1};s\right)-\frac{5}{4} J\left(\Omega _0,\omega
_4^{z_1},\zeta _{4,-1}^{z_1};s\right)\\ -\frac{5}{4} J\left(\Omega _0,\omega _4^{z_1},\zeta _{4,1}^{z_1};s\right)+\frac{1}{6} J\left(\Omega _1,\omega
_{-1}^{\text{sz}},\omega _0^z;s\right)+\frac{1}{6} J\left(\Omega _1,\omega _1^{\text{sz}},\omega _0^z;s\right)-\frac{5}{4} J\left(\Omega _1,\omega
_1^{z_1},\omega _{4,-1}^{z_1};s\right)\\ -\frac{5}{4} J\left(\Omega _1,\omega _1^{z_1},\omega _{4,1}^{z_1};s\right)+\frac{5}{4} J\left(\Omega _1,\omega
_1^{z_1},\zeta _{4,-1}^{z_1};s\right)-\frac{5}{4} J\left(\Omega _1,\omega _1^{z_1},\zeta _{4,1}^{z_1};s\right)+\frac{5}{4} J\left(\Omega _1,\omega
_4^{z_1},\omega _{4,-1}^{z_1};s\right) \\ +\frac{5}{4} J\left(\Omega _1,\omega _4^{z_1},\omega _{4,1}^{z_1};s\right)-\frac{5}{4} J\left(\Omega _1,\omega
_4^{z_1},\zeta _{4,-1}^{z_1};s\right)+\frac{5}{4} J\left(\Omega _1,\omega _4^{z_1},\zeta _{4,1}^{z_1};s\right)-\frac{5}{6} J\left(\Omega _1,\omega
_{1,-1}^s,\omega _0^z;s\right)\\ -\frac{5}{6} J\left(\Omega _1,\omega _{1,1}^s,\omega _0^z;s\right)+J\left(\Omega _0,\omega _1^s,\omega _4^{z_1},\omega
_4^{z_1};s\right)+\frac{5}{4} J\left(\Omega _0,\omega _1^{z_1},\eta _{4,-1}^{z_1},\omega _0^z;s\right)-\frac{5}{4} J\left(\Omega _0,\omega _1^{z_1},\eta
_{4,1}^{z_1},\omega _0^z;s\right)\\ +\frac{5}{8} J\left(\Omega _0,\omega _1^{z_1},\omega _{4,-1}^{z_1},\omega _0^z;s\right)-\frac{5}{8} J\left(\Omega _0,\omega
_1^{z_1},\omega _{4,1}^{z_1},\omega _0^z;s\right)-\frac{15}{8} J\left(\Omega _0,\omega _1^{z_1},\zeta _{4,-1}^{z_1},\omega _0^z;s\right)\\ -\frac{15}{8}
J\left(\Omega _0,\omega _1^{z_1},\zeta _{4,1}^{z_1},\omega _0^z;s\right) -\frac{5}{4} J\left(\Omega _0,\omega _4^{z_1},\eta _{4,-1}^{z_1},\omega
_0^z;s\right)+\frac{5}{4} J\left(\Omega _0,\omega _4^{z_1},\eta _{4,1}^{z_1},\omega _0^z;s\right)-J\left(\Omega _0,\omega _4^{z_1},\omega _1^s,\omega
_4^{z_1};s\right)\\ -J\left(\Omega _0,\omega _4^{z_1},\omega _4^{z_1},\omega _1^s;s\right) -\frac{5}{8} J\left(\Omega _0,\omega _4^{z_1},\omega _{4,-1}^{z_1},\omega
_0^z;s\right)+\frac{5}{8} J\left(\Omega _0,\omega _4^{z_1},\omega _{4,1}^{z_1},\omega _0^z;s\right)+\frac{15}{8} J\left(\Omega _0,\omega _4^{z_1},\zeta
_{4,-1}^{z_1},\omega _0^z;s\right)\\ +\frac{15}{8} J\left(\Omega _0,\omega _4^{z_1},\zeta _{4,1}^{z_1},\omega _0^z;s\right) +\frac{5}{4} J\left(\Omega _1,\omega
_1^{z_1},\eta _{4,-1}^{z_1},\omega _0^z;s\right)+\frac{5}{4} J\left(\Omega _1,\omega _1^{z_1},\eta _{4,1}^{z_1},\omega _0^z;s\right)\\ +\frac{1}{2} J\left(\Omega
_1,\omega _1^{z_1},\omega _{4,-1}^{z_1},\omega _0^z;s\right)+\frac{1}{2} J\left(\Omega _1,\omega _1^{z_1},\omega _{4,1}^{z_1},\omega _0^z;s\right)-\frac{15}{8}
J\left(\Omega _1,\omega _1^{z_1},\zeta _{4,-1}^{z_1},\omega _0^z;s\right)\\ +\frac{15}{8} J\left(\Omega _1,\omega _1^{z_1},\zeta _{4,1}^{z_1},\omega
_0^z;s\right) -\frac{5}{4} J\left(\Omega _1,\omega _4^{z_1},\eta _{4,-1}^{z_1},\omega _0^z;s\right)-\frac{5}{4} J\left(\Omega _1,\omega _4^{z_1},\eta
_{4,1}^{z_1},\omega _0^z;s\right)\\ -\frac{1}{2} J\left(\Omega _1,\omega _4^{z_1},\omega _{4,-1}^{z_1},\omega _0^z;s\right)-\frac{1}{2} J\left(\Omega _1,\omega
_4^{z_1},\omega _{4,1}^{z_1},\omega _0^z;s\right)+\frac{15}{8} J\left(\Omega _1,\omega _4^{z_1},\zeta _{4,-1}^{z_1},\omega _0^z;s\right) \\ -\frac{15}{8}
J\left(\Omega _1,\omega _4^{z_1},\zeta _{4,1}^{z_1},\omega _0^z;s\right)
\Big]  + \OO(\ep)
\end{multline}
and 
\begin{multline}
e^{2\ep\gamma_E}j_2^{\kite}(0,1,1,1,1) =	
\frac{1}{2 \ep ^2}+\frac{z_1 J\left(\Omega _0,\omega _4^{z_1};s\right)+\frac{5}{2}}{\ep }+\frac{3 \left(s^2-10 s+9\right) J\left(\Theta _8,\omega _0^z;s\right)}{4 s}\\ + \left(-\frac{s}{2}-\frac{9}{2 s}+5\right) J\left(\Theta _5;s\right)+\frac{1}{2}
\left(s+\frac{9}{s}-10\right) J\left(\Theta _6;s\right)+\left(-\frac{s}{2}-\frac{9}{2 s}+5\right) J\left(\Theta _8;s\right)\\ +\frac{1}{4}
\left(s+\frac{9}{s}-10\right) J\left(\Theta _5,\omega _0^z;s\right)+\left(-\frac{s}{3}-\frac{3}{s}+\frac{5}{6}\right) J\left(\Theta _6,\omega
_0^z;s\right)+\left(-\frac{s}{2}-\frac{9}{2 s}+5\right) J\left(\Theta _9,\omega _0^z;s\right)\\ -\frac{45}{16} z_1 J\left(\Omega _0,\zeta _{4,-1}^{z_1},\omega
_0^z;s\right)-\frac{45}{16} z_1 J\left(\Omega _0,\zeta _{4,1}^{z_1},\omega _0^z;s\right)-\frac{45}{16} z_1 J\left(\Omega _1,\zeta
_{4,-1}^{z_1},\omega _0^z;s\right)+\frac{45}{16} z_1 J\left(\Omega _1,\zeta _{4,1}^{z_1},\omega _0^z;s\right)\\ +\frac{15}{8} z_1 J\left(\Omega _0,\zeta
_{4,-1}^{z_1};s\right)+\frac{15}{8} z_1 J\left(\Omega _0,\zeta _{4,1}^{z_1};s\right)+\frac{15}{8} z_1 J\left(\Omega _1,\zeta
_{4,-1}^{z_1};s\right)-\frac{15}{8} z_1 J\left(\Omega _1,\zeta _{4,1}^{z_1};s\right) \\ +\frac{15}{8} z_1 J\left(\Omega _0,\eta _{4,-1}^{z_1},\omega
_0^z;s\right)-\frac{15}{8} z_1 J\left(\Omega _0,\eta _{4,1}^{z_1},\omega _0^z;s\right)+\frac{15}{8} z_1 J\left(\Omega _1,\eta _{4,-1}^{z_1},\omega
_0^z;s\right)+\frac{15}{8} z_1 J\left(\Omega _1,\eta _{4,1}^{z_1},\omega _0^z;s\right)\\ +4 z_1 J\left(\Omega _0,\omega _4^{z_1};s\right)-\frac{15}{8}
z_1 J\left(\Omega _0,\omega _{4,-1}^{z_1};s\right)+\frac{15}{8} z_1 J\left(\Omega _0,\omega _{4,1}^{z_1};s\right)-\frac{15}{8} z_1
J\left(\Omega _1,\omega _{4,-1}^{z_1};s\right)\\-\frac{15}{8} z_1 J\left(\Omega _1,\omega _{4,1}^{z_1};s\right)-z_1 J\left(\Omega _0,\omega _4^s,\omega
_4^{z_1};s\right)+\frac{15}{16} z_1 J\left(\Omega _0,\omega _{4,-1}^{z_1},\omega _0^z;s\right)-\frac{15}{16} z_1 J\left(\Omega _0,\omega
_{4,1}^{z_1},\omega _0^z;s\right)\\ +\frac{3}{4} z_1 J\left(\Omega _1,\omega _{4,-1}^{z_1},\omega _0^z;s\right)+\frac{3}{4} z_1 J\left(\Omega _1,\omega
_{4,1}^{z_1},\omega _0^z;s\right)+\frac{1}{12} \left(-81 S_2+\pi^2+114\right) + \OO(\ep)
\end{multline} 
The results for other master integrals together with ${\cal O}(\ep^1)$ terms for $j_2^{\kite}(1,1,1,1,1)$ and $j_2^{\kite}(0,1,1,1,1)$ integrals can be found in accompanying Mathematica file. Here we would like to stress that already in the case of kite with only one massless line the results, because of additional square-root $z_1$ in addition to one present\footnote{See Appendix \ref{notation} for notation.} in $J_y$,  can not be written in terms of elliptic polylogarithms and one needs to enlarge this class of functions to include hyperelliptic polylogarithms, related to algebraic curves of degree higher than one. On the other hand our class of functions defined as iterated integrals with algebraic kernels is fully sufficient. 

\subsection{Fully massive kite diagram}
\begin{figure}[h]
	\centering
	\includegraphics[scale=0.6]{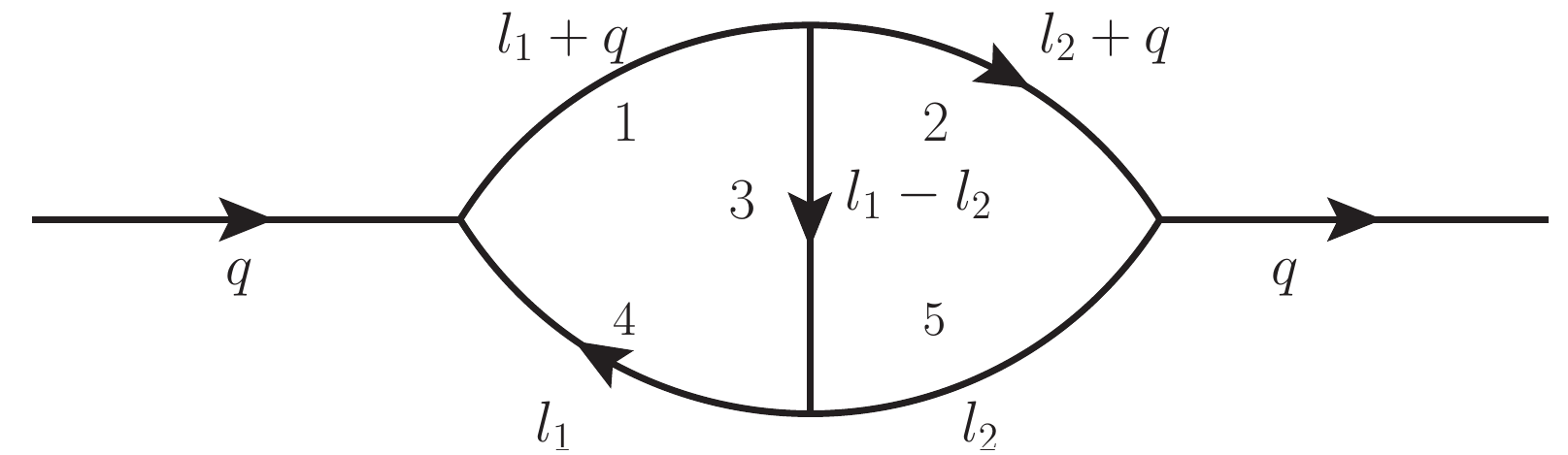}
	\caption{Fully massive kite diagram.}
	\label{fig:kite3}
\end{figure}

Finally, let us discuss the last case of fully massive kite diagram depicted in Fig. \ref{fig:kite3}. The master integrals in this family are defined as ($d=4-2\ep$):
\begin{equation}
j_3^{\kite}(a_1,\ldots , a_5) = \int\frac{d^d l_1 d^d l_2}{(i\pi^{d/2})^2}
\frac{1}{D_1^{a_1}D_2^{a_2}D_3^{a_3}D_4^{a_4}D_5^{a_5}},
\end{equation}
where
\begin{align}
D_1 &= 1-(l_1+q)^2,& D_2 &= 1-(l_2+q)^2,& D_3 &= 1-(l_1-l_2)^2,& D_4 &= 1-l_4^2,& D_5 &= 1-l_2^2\, . \nonumber \\
\end{align}
and the vector of Laporta master integrals can be chosen as: 
\begin{align}
I_{\rm Laporta} = \{
&j_3^{\kite}(0,0,0,1,1),& &j_3^{\kite}(0,0,1,1,1),& &j_3^{\kite}(0,1,0,1,1),& &S_{1,1,1}^{(2)},\nonumber \\  &S_{2,1,1}^{(2)},& &j_3^{\kite}(0,1,1,1,1),& &j_3^{\kite}(1,1,0,1,1),& &j_3^{\kite}(1,1,1,1,1)
\}^{\top}
\end{align}

\noindent
Reducing similar to two previous cases the system of differential equations with respect to $q^2=s$ to $A+B\ep$ form we get ($I_{\rm Laporta}=T\tilde{I}_{\rm canonical}$):
\begin{equation}
\frac{d\tilde{I}_{\rm canonical}}{ds} = M\tilde{I}_{\rm canonical}\, ,
\end{equation}
where ($z_1 = \sqrt{\frac{4-s}{s}}$):
\begin{equation}
M = \frac{M_1}{s-9} + \frac{M_2}{s-4} + \frac{M_3}{s-3} + \frac{M_4}{s-1} + \frac{M_5}{s}
+ \frac{z_1 M_6}{s-9} + \frac{z_1 M_7}{s-4} + \frac{z_1 M_8}{s-3}
\end{equation}
Again, the particular expressions for coefficient matrices $M_i$ together with transformation matrix to canonical basis $T$ can be found in accompanying Mathematica notebook. The solution of differential system in the canonical basis goes similar to the two cases considered previously and for $j_3^{\kite}(1,1,1,1,1)$ master we get\footnote{The only other master with elliptics different from sunsets was already evaluated in previous subsection $j_3^{\kite}(0,1,1,1,1) = j_2^{\kite}(0,1,1,1,1)$}:
\begin{multline}
e^{2\ep\gamma_E}j_3^{\kite}(1,1,1,1,1) = \frac{1}{s}\Big[-\frac{15}{2} J\left(\Omega _0,\omega _3^{z_1},\omega _{4,-1}^{z_1};s\right)+\frac{15}{2} J\left(\Omega _0,\omega _3^{z_1},\omega _{4,1}^{z_1};s\right) \\ +\frac{15}{2}
J\left(\Omega _0,\omega _3^{z_1},\zeta _{4,-1}^{z_1};s\right)  +\frac{15}{2} J\left(\Omega _0,\omega _3^{z_1},\zeta _{4,1}^{z_1};s\right)+\frac{15}{2}
J\left(\Omega _0,\omega _4^{z_1},\omega _{4,-1}^{z_1};s\right)\\-\frac{15}{2} J\left(\Omega _0,\omega _4^{z_1},\omega _{4,1}^{z_1};s\right)-\frac{15}{2}
J\left(\Omega _0,\omega _4^{z_1},\zeta _{4,-1}^{z_1};s\right)-\frac{15}{2} J\left(\Omega _0,\omega _4^{z_1},\zeta _{4,1}^{z_1};s\right) \\ +\frac{1}{3} J\left(\Omega
_1,\omega _{-1}^{\text{sz}},\omega _0^z;s\right)+\frac{1}{3} J\left(\Omega _1,\omega _1^{\text{sz}},\omega _0^z;s\right)-\frac{15}{2} J\left(\Omega _1,\omega
_3^{z_1},\omega _{4,-1}^{z_1};s\right)\\-\frac{15}{2} J\left(\Omega _1,\omega _3^{z_1},\omega _{4,1}^{z_1};s\right)+\frac{15}{2} J\left(\Omega _1,\omega
_3^{z_1},\zeta _{4,-1}^{z_1};s\right)-\frac{15}{2} J\left(\Omega _1,\omega _3^{z_1},\zeta _{4,1}^{z_1};s\right) \\ +\frac{15}{2} J\left(\Omega _1,\omega
_4^{z_1},\omega _{4,-1}^{z_1};s\right)+\frac{15}{2} J\left(\Omega _1,\omega _4^{z_1},\omega _{4,1}^{z_1};s\right)-\frac{15}{2} J\left(\Omega _1,\omega
_4^{z_1},\zeta _{4,-1}^{z_1};s\right)\\ +\frac{15}{2} J\left(\Omega _1,\omega _4^{z_1},\zeta _{4,1}^{z_1};s\right)-J\left(\Omega _1,\omega _{3,-1}^s,\omega
_0^z;s\right)-J\left(\Omega _1,\omega _{3,1}^s,\omega _0^z;s\right) \\ -2 J\left(\Omega _0,\omega _3^s,\omega _4^{z_1},\omega _4^{z_1};s\right)+\frac{15}{2}
J\left(\Omega _0,\omega _3^{z_1},\eta _{4,-1}^{z_1},\omega _0^z;s\right)-\frac{15}{2} J\left(\Omega _0,\omega _3^{z_1},\eta _{4,1}^{z_1},\omega
_0^z;s\right)\\+\frac{15}{4} J\left(\Omega _0,\omega _3^{z_1},\omega _{4,-1}^{z_1},\omega _0^z;s\right)-\frac{15}{4} J\left(\Omega _0,\omega _3^{z_1},\omega
_{4,1}^{z_1},\omega _0^z;s\right)-\frac{45}{4} J\left(\Omega _0,\omega _3^{z_1},\zeta _{4,-1}^{z_1},\omega _0^z;s\right)\\-\frac{45}{4} J\left(\Omega _0,\omega
_3^{z_1},\zeta _{4,1}^{z_1},\omega _0^z;s\right)-\frac{15}{2} J\left(\Omega _0,\omega _4^{z_1},\eta _{4,-1}^{z_1},\omega _0^z;s\right)+\frac{15}{2} J\left(\Omega
_0,\omega _4^{z_1},\eta _{4,1}^{z_1},\omega _0^z;s\right)\\-\frac{15}{4} J\left(\Omega _0,\omega _4^{z_1},\omega _{4,-1}^{z_1},\omega _0^z;s\right)+\frac{15}{4}
J\left(\Omega _0,\omega _4^{z_1},\omega _{4,1}^{z_1},\omega _0^z;s\right)+\frac{45}{4} J\left(\Omega _0,\omega _4^{z_1},\zeta _{4,-1}^{z_1},\omega
_0^z;s\right)\\+\frac{45}{4} J\left(\Omega _0,\omega _4^{z_1},\zeta _{4,1}^{z_1},\omega _0^z;s\right)+\frac{15}{2} J\left(\Omega _1,\omega _3^{z_1},\eta
_{4,-1}^{z_1},\omega _0^z;s\right)+\frac{15}{2} J\left(\Omega _1,\omega _3^{z_1},\eta _{4,1}^{z_1},\omega _0^z;s\right)\\+3 J\left(\Omega _1,\omega _3^{z_1},\omega
_{4,-1}^{z_1},\omega _0^z;s\right)+3 J\left(\Omega _1,\omega _3^{z_1},\omega _{4,1}^{z_1},\omega _0^z;s\right)-\frac{45}{4} J\left(\Omega _1,\omega
_3^{z_1},\zeta _{4,-1}^{z_1},\omega _0^z;s\right)\\+\frac{45}{4} J\left(\Omega _1,\omega _3^{z_1},\zeta _{4,1}^{z_1},\omega _0^z;s\right)-\frac{15}{2}
J\left(\Omega _1,\omega _4^{z_1},\eta _{4,-1}^{z_1},\omega _0^z;s\right)-\frac{15}{2} J\left(\Omega _1,\omega _4^{z_1},\eta _{4,1}^{z_1},\omega _0^z;s\right)\\-3
J\left(\Omega _1,\omega _4^{z_1},\omega _{4,-1}^{z_1},\omega _0^z;s\right)-3 J\left(\Omega _1,\omega _4^{z_1},\omega _{4,1}^{z_1},\omega
_0^z;s\right)+\frac{45}{4} J\left(\Omega _1,\omega _4^{z_1},\zeta _{4,-1}^{z_1},\omega _0^z;s\right)\\-\frac{45}{4} J\left(\Omega _1,\omega _4^{z_1},\zeta
_{4,1}^{z_1},\omega _0^z;s\right)
\Big] + \OO(\ep).
\end{multline}
The results for other master integrals together with ${\cal O}(\ep^1)$ term for $j_3^{\kite}(1,1,1,1,1)$  master can be found in accompanying Mathematica file. Here we again facing a situation when the result can not be written in terms of elliptic polylogarithms and one needs to introduce hyperelliptic polylogarithms. On the other hand the class of functions defined as iterated integrals with algebraic kernels is fully sufficient. 

\section{Conclusion}		

Using elliptic kite master integrals as example we have shown, that class of functions,
defined as iterated integrals with algebraic kernels\footnote{See Appendix A.},
is fully sufficient to express their $\ep$-expansion terms to all orders in $\ep$.
Moreover, the obtained results for kite master integral with two massless lines are more compact,
than those available in literature.
The analytical results for two other kite diagrams are new and where not known before.
To make such presentation possible we used the new integral representations
for sunset subgraphs in $d=4-2\ep$ and $d=2-2\ep$ dimensions together with differential equations for considered kite master integrals in $A+B\ep$ form.
This form is expected to be applicable for most if not all differential systems, related to Feynman diagrams.
The obtained sunset integral representations can be used further to get results for other elliptic master integrals with elliptic sunset subgraphs. We have checked,
using sector decomposition method \cite{Binoth:2000ps,Binoth:2003ak,Binoth:2004jv,Heinrich:2008si,Bogner:2007cr,Bogner:2008ry,Kaneko:2009qx} as implemented in \cite{Fiesta4}, that the obtained sunset integral representations are well defined both below and above thresholds.
The results for kite diagrams where checked only below threshold
and their analytical continuation above it will be the subject of one of our future publications.


We would like to thank R.N.Lee for interesting and stimulating discussions.
The work of M.A.B and A.I.O was supported in part by the Foundation for the
Advancement of Theoretical Physics and Mathematics “BASIS” and Russian Science Foundation, grant 20-12-00205.
The work of O.L.V. was supported in part by DFG Research Unit FOR 2926 through Grant No. 409651613.
The authors also would like to thank Heisenberg-Landau program.

\appendix

\section{Notation for iterated integrals}\label{notation}

The $G$-functions one may encounter along this paper stand for the usual multiple polylogarithms \cite{polylog1,polylog2,polylog3} and are defined as the following iterated integrals 
\begin{equation}
G(a_1,\ldots ,a_k; y) = \int_0^y \frac{dt}{t-a_1}G(a_2,\ldots , a_k; t), \quad G(\underbrace{0,\ldots ,0}_{k}; y) = \frac{1}{k!}(\ln y)^k
\end{equation}
The obtained in the paper results for sunset and kite diagrams can be conveniently expressed in terms of iterated integrals with algebraic kernels of the form:
\begin{equation}
J(\Omega,\omega_1^s,\ldots ,\omega_n^s,\omega_1^z,\ldots ,\omega_m^z,\omega_1^y,\ldots ,\omega_l^y;s)
\end{equation}
where $\Omega$ is some 1-form in $y$ integrated from 2 to $\infty$, for example $\frac{dy}{y^2\sqrt{y^2-4}}$. $\omega_n^s$ are some 1-forms in $s$, for example $\frac{ds}{s-4}$. $\omega_m^z$ - some 1-forms in $z$ $(z(s,y) = \frac{1-s+y^2+\sqrt{(s-1-y^2)^2-4y^2}}{2y})$, for example $\frac{dz}{z-y}$. Finally, $\omega_l^y$ are some 1-forms in $y$, for example $\frac{dy}{y}$. Integrals with respect to $\omega_n^s$, $\omega_m^z$, $\omega_l^y$ forms are iterated integrals in $s$, $z$ and $y$ correspondingly. For example, we have
\begin{equation}
J\left(
\Omega , \frac{ds}{s-4}, \frac{dz}{z-y}, \frac{dy}{y}; s
\right) = \int_2^{\infty}\frac{dy}{y^2\sqrt{y^2-4}}\int_0^s\frac{ds'}{s'-4}\int_0^{z(s',y)}\frac{dz'}{z'-y}\int_0^y\frac{dy'}{y'}\, .
\end{equation}
In general, iterated integrals in our results contain the following $\Omega$ 1-forms ($J_y=\frac{4}{y^2\sqrt{y^2-4}}$): 
\begin{align}
\Omega_0 &= J_y dy, &  \Omega_1 &= y J_y dy, & \Omega_2 &= \frac{J_y dy}{y}, & \Omega_3 &= \frac{J_y dy}{y-1}, & \Omega_4 &= \frac{J_y dy}{y+1}, 
\end{align}
$\Theta$ 1-forms:
\begin{align}
\Theta_y &= \frac{y J_y dy}{z-y}, & \Theta_{1/y} &= \frac{J_y dy}{zy-1}, & \Theta_1 &= \frac{y J_y dy}{(y^2-1)(z-y)},  \nonumber \\  \Theta_2 &= \frac{J_y dy}{(y^2-1)(zy-1)}, &
\Theta_3 &= \frac{y J_y dy}{(y-1)(z-1)}, & \Theta_4 &= \frac{y J_y dy}{(y+1)(z+1)}, \nonumber \\ 
\Theta_5 &= \frac{J_y dy}{z^2-1}, & \Theta_6 &= \frac{yz J_y dy}{z^2-1}, & 
\Theta_7 &= 
\frac{J_y dy}{2(z+1)^2} + \frac{J_y dy}{2(z-1)^2}, \nonumber \\
\Theta_8 &= 
\frac{(1+y)J_ydy}{2(z+1)^2} + \frac{(1-y)J_y dy}{2(z-1)^2}
, &   \Theta_9 &= \frac{(1+y)J_ydy}{2(z+1)^3} + \frac{(y-1)J_y dy}{2(z-1)^3}\, 
\end{align}
and  $\omega$, $\zeta$ and $\eta$ 1-forms ($z_1 = \sqrt{\frac{4-s}{s}}$):
\begin{align}
\omega_a^y &= \frac{dy}{y-a}, & \omega_a^z &= \frac{dz}{z-a}, & \omega_a^s &= \frac{ds}{s-a}, & \omega_a^{\text{sz}} &= \frac{ds}{z-a}, \nonumber \\ 
\omega_a^{z_1} &= \frac{z_1 ds}{s-a}, & \omega_{a,b}^s &= \frac{ds}{(s-a)(z-b)}, &
\omega_{a,b}^{z_1} &= \frac{z_1 ds}{(s-a)(z-b)}, & \zeta_{a,b}^s &= \frac{ds}{(s-a)(z-b)^2}, \nonumber \\ 
\zeta_{a,b}^{z_1} &= \frac{z_1 ds}{(s-a)(z-b)^2}, & \eta_{a,b}^s &= \frac{ds}{(s-a)(z-b)^3}, & \eta_{a,b}^{z_1} &= \frac{z_1 ds}{(s-a)(z-b)^3}\, .
\end{align}
Note, that integrals over $s$ variables can be easily rewritten in terms of integrals over $z$ variables
and actually we have only 1-forms in $y$ and $z$. The use of 1-forms in $s$ simply makes their expressions more compact.

\section{Results for kite masters with $d=4-2\ep$ sunsets}

The evaluation of kite master integrals in the case of $d=4-2\ep$ sunset subgraphs goes along the same lines as in the case of $d=2-2\ep$ sunsets considered in the main body of the paper. In particular we have\footnote{All other details and results may be found in accompanying Mathematica notebook.}
\small
\begin{multline}
e^{2\ep\gamma_E}j_1^{\kite}(1,1,1,1,1) = \frac{1}{6s}\Big[\left(-16+\frac{5 \pi ^2}{3}\right) G(1;s)+\left(4-\frac{\pi ^2}{6}\right) G(7;s)+12 G(0,1,1;s) -6 G(1,0,1;s) \\ +8 J\left(\Omega _0,\omega _{1,-1}^s,\omega
_0^z;s\right)-8 J\left(\Omega _0,\omega _{1,1}^s,\omega _0^z;s\right)-2 J\left(\Omega _0,\omega _{7,-1}^s,\omega _0^z;s\right)  +2 J\left(\Omega _0,\omega
_{7,1}^s,\omega _0^z;s\right) -8 J\left(\Omega _1,\omega _{1,y}^s,\omega _0^y;s\right) \\ +8 J\left(\Omega _1,\omega _{1,y}^s,\omega _0^z;s\right)+2 J\left(\Omega
_1,\omega _{7,y}^s,\omega _0^y;s\right)  -2 J\left(\Omega _1,\omega _{7,y}^s,\omega _0^z;s\right)  -8 J\left(\Omega _3,\omega _{1,1}^s,\omega _0^z;s\right)-4
J\left(\Omega _3,\omega _{1,y}^s,\omega _0^y;s\right) \\ +4 J\left(\Omega _3,\omega _{1,y}^s,\omega _0^z;s\right)  +4 J\left(\Omega _3,\omega _{1,\bar{y}}^s,\omega
_0^y;s\right)  +4 J\left(\Omega _3,\omega _{1,\bar{y}}^s,\omega _0^z;s\right)+2 J\left(\Omega _3,\omega _{7,1}^s,\omega _0^z;s\right)+J\left(\Omega _3,\omega
_{7,y}^s,\omega _0^y;s\right) \\ -J\left(\Omega _3,\omega _{7,y}^s,\omega _0^z;s\right)-J\left(\Omega _3,\omega _{7,\bar{y}}^s,\omega _0^y;s\right)-J\left(\Omega
_3,\omega _{7,\bar{y}}^s,\omega _0^z;s\right)-8 J\left(\Omega _4,\omega _{1,-1}^s,\omega _0^z;s\right)  -4 J\left(\Omega _4,\omega _{1,y}^s,\omega _0^y;s\right) \\ +4
J\left(\Omega _4,\omega _{1,y}^s,\omega _0^z;s\right)+4 J\left(\Omega _4,\omega _{1,\bar{y}}^s,\omega _0^y;s\right)+4 J\left(\Omega _4,\omega
_{1,\bar{y}}^s,\omega _0^z;s\right)  +2 J\left(\Omega _4,\omega _{7,-1}^s,\omega _0^z;s\right)+J\left(\Omega _4,\omega _{7,y}^s,\omega _0^y;s\right) \\ -J\left(\Omega
_4,\omega _{7,y}^s,\omega _0^z;s\right)-J\left(\Omega _4,\omega _{7,\bar{y}}^s,\omega _0^y;s\right)  -J\left(\Omega _4,\omega _{7,\bar{y}}^s,\omega _0^z;s\right)+8
J\left(\Omega _0,\omega _1^s,\omega _0^y,\omega _0^y;s\right)+8 J\left(\Omega _0,\omega _1^s,\omega _0^z,\omega _0^y;s\right) \\ -8 J\left(\Omega _0,\omega
_1^s,\omega _0^z,\omega _0^z;s\right)  -8 J\left(\Omega _0,\omega _1^s,\omega _y^z,\omega _0^y;s\right)+8 J\left(\Omega _0,\omega _1^s,\omega _y^z,\omega
_0^z;s\right)-8 J\left(\Omega _0,\omega _1^s,\omega _{\bar{y}}^z,\omega _0^y;s\right) \\-8 J\left(\Omega _0,\omega _1^s,\omega _{\bar{y}}^z,\omega _0^z;s\right) -2
J\left(\Omega _0,\omega _7^s,\omega _0^y,\omega _0^y;s\right)-2 J\left(\Omega _0,\omega _7^s,\omega _0^z,\omega _0^y;s\right)+2 J\left(\Omega _0,\omega
_7^s,\omega _0^z,\omega _0^z;s\right) \\ +2 J\left(\Omega _0,\omega _7^s,\omega _y^z,\omega _0^y;s\right) -2 J\left(\Omega _0,\omega _7^s,\omega _y^z,\omega
_0^z;s\right)+2 J\left(\Omega _0,\omega _7^s,\omega _{\bar{y}}^z,\omega _0^y;s\right)+2 J\left(\Omega _0,\omega _7^s,\omega _{\bar{y}}^z,\omega _0^z;s\right)\Big] + \OO(\ep).
\end{multline}

\begin{multline}
e^{2\ep\gamma_E}j_2^{\kite}(1,1,1,1,1) =\frac{1}{s}\Big[\frac{5}{36} \left(-24+\pi ^2\right) G(1;s)+\left(\frac{4}{3}-\frac{\pi ^2}{18}\right) G(7;s)  +\frac{5}{144} \left(-12+\pi ^2\right) J\left(\Omega _0,\omega
_1^{z_1},\omega _1^{z_1};s\right) \\ -\frac{4}{27} \left(-15+\pi ^2\right) J\left(\Omega _0,\omega _1^{z_1},\omega _4^{z_1};s\right)  +\frac{19}{108} \left(-24+\pi
^2\right) J\left(\Omega _0,\omega _1^{z_1},\omega _7^{z_1};s\right)  +\left(\frac{3}{4}-\frac{\pi ^2}{16}\right) J\left(\Omega _0,\omega _1^{z_1},\omega_9^{z_1};s\right) \\-\frac{5}{144} \left(-12+\pi ^2\right) J\left(\Omega _0,\omega _4^{z_1},\omega _1^{z_1};s\right)+\frac{4}{27} \left(-15+\pi ^2\right)
J\left(\Omega _0,\omega _4^{z_1},\omega _4^{z_1};s\right) -\frac{19}{108} \left(-24+\pi ^2\right) J\left(\Omega _0,\omega _4^{z_1},\omega
_7^{z_1};s\right) \\ +\frac{1}{16} \left(-12+\pi ^2\right) J\left(\Omega _0,\omega _4^{z_1},\omega _9^{z_1};s\right)  +\frac{5}{3} J\left(\Omega _0,\omega
_{1,-1}^s,\omega _0^z;s\right)-\frac{5}{3} J\left(\Omega _0,\omega _{1,1}^s,\omega _0^z;s\right)-\frac{2}{3} J\left(\Omega _0,\omega _{7,-1}^s,\omega
_0^z;s\right) \\ +\frac{2}{3} J\left(\Omega _0,\omega _{7,1}^s,\omega _0^z;s\right)  -\frac{5}{3} J\left(\Omega _1,\omega _{1,y}^s,\omega _0^y;s\right)+\frac{5}{3}
J\left(\Omega _1,\omega _{1,y}^s,\omega _0^z;s\right)+\frac{2}{3} J\left(\Omega _1,\omega _{7,y}^s,\omega _0^y;s\right)-\frac{2}{3} J\left(\Omega _1,\omega
_{7,y}^s,\omega _0^z;s\right) \\ -\frac{5}{3} J\left(\Omega _3,\omega _{1,1}^s,\omega _0^z;s\right)-\frac{5}{6} J\left(\Omega _3,\omega _{1,y}^s,\omega
_0^y;s\right)+\frac{5}{6} J\left(\Omega _3,\omega _{1,y}^s,\omega _0^z;s\right)+\frac{5}{6} J\left(\Omega _3,\omega _{1,\bar{y}}^s,\omega
_0^y;s\right)  +\frac{5}{6} J\left(\Omega _3,\omega _{1,\bar{y}}^s,\omega _0^z;s\right) \\ +\frac{2}{3} J\left(\Omega _3,\omega _{7,1}^s,\omega
_0^z;s\right)+\frac{1}{3} J\left(\Omega _3,\omega _{7,y}^s,\omega _0^y;s\right)-\frac{1}{3} J\left(\Omega _3,\omega _{7,y}^s,\omega _0^z;s\right)  -\frac{1}{3}
J\left(\Omega _3,\omega _{7,\bar{y}}^s,\omega _0^y;s\right)  -\frac{1}{3} J\left(\Omega _3,\omega _{7,\bar{y}}^s,\omega _0^z;s\right) \\ -\frac{5}{3} J\left(\Omega
_4,\omega _{1,-1}^s,\omega _0^z;s\right)-\frac{5}{6} J\left(\Omega _4,\omega _{1,y}^s,\omega _0^y;s\right)  +\frac{5}{6} J\left(\Omega _4,\omega _{1,y}^s,\omega
_0^z;s\right)+\frac{5}{6} J\left(\Omega _4,\omega _{1,\bar{y}}^s,\omega _0^y;s\right)+\frac{5}{6} J\left(\Omega _4,\omega _{1,\bar{y}}^s,\omega
_0^z;s\right) \\ +\frac{2}{3} J\left(\Omega _4,\omega _{7,-1}^s,\omega _0^z;s\right)  +\frac{1}{3} J\left(\Omega _4,\omega _{7,y}^s,\omega _0^y;s\right)-\frac{1}{3}
J\left(\Omega _4,\omega _{7,y}^s,\omega _0^z;s\right)-\frac{1}{3} J\left(\Omega _4,\omega _{7,\bar{y}}^s,\omega _0^y;s\right)-\frac{1}{3} J\left(\Omega _4,\omega
_{7,\bar{y}}^s,\omega _0^z;s\right) \\ +\frac{5}{3} J\left(\Omega _0,\omega _1^s,\omega _0^y,\omega _0^y;s\right)+\frac{5}{3} J\left(\Omega _0,\omega _1^s,\omega
_0^z,\omega _0^y;s\right)-\frac{5}{3} J\left(\Omega _0,\omega _1^s,\omega _0^z,\omega _0^z;s\right)+J\left(\Omega _0,\omega _1^s,\omega _4^{z_1},\omega
_4^{z_1};s\right) \\ -\frac{5}{3} J\left(\Omega _0,\omega _1^s,\omega _y^z,\omega _0^y;s\right)+\frac{5}{3} J\left(\Omega _0,\omega _1^s,\omega _y^z,\omega
_0^z;s\right)-\frac{5}{3} J\left(\Omega _0,\omega _1^s,\omega _{\bar{y}}^z,\omega _0^y;s\right)-\frac{5}{3} J\left(\Omega _0,\omega _1^s,\omega
_{\bar{y}}^z,\omega _0^z;s\right) \\ +\frac{5}{12} J\left(\Omega _0,\omega _1^{z_1},\omega _1^{z_1},\omega _0^z;s\right)-\frac{4}{3} J\left(\Omega _0,\omega
_1^{z_1},\omega _4^{z_1},\omega _0^z;s\right)-\frac{3}{4} J\left(\Omega _0,\omega _1^{z_1},\omega _9^{z_1},\omega _0^z;s\right)-\frac{4}{9} J\left(\Omega
_0,\omega _1^{z_1},\omega _{4,-1}^{z_1},\omega _0^z;s\right) \\ +\frac{4}{9} J\left(\Omega _0,\omega _1^{z_1},\omega _{4,1}^{z_1},\omega _0^z;s\right)+\frac{19}{9}
J\left(\Omega _0,\omega _1^{z_1},\omega _{7,-1}^{z_1},\omega _0^z;s\right)-\frac{19}{9} J\left(\Omega _0,\omega _1^{z_1},\omega _{7,1}^{z_1},\omega
_0^z;s\right)-J\left(\Omega _0,\omega _4^{z_1},\omega _1^s,\omega _4^{z_1};s\right) \\ -\frac{5}{12} J\left(\Omega _0,\omega _4^{z_1},\omega _1^{z_1},\omega
_0^z;s\right)+\frac{4}{3} J\left(\Omega _0,\omega _4^{z_1},\omega _4^{z_1},\omega _0^z;s\right)-J\left(\Omega _0,\omega _4^{z_1},\omega _4^{z_1},\omega
_1^s;s\right)+\frac{3}{4} J\left(\Omega _0,\omega _4^{z_1},\omega _9^{z_1},\omega _0^z;s\right) \\ +\frac{4}{9} J\left(\Omega _0,\omega _4^{z_1},\omega
_{4,-1}^{z_1},\omega _0^z;s\right)-\frac{4}{9} J\left(\Omega _0,\omega _4^{z_1},\omega _{4,1}^{z_1},\omega _0^z;s\right)-\frac{19}{9} J\left(\Omega _0,\omega
_4^{z_1},\omega _{7,-1}^{z_1},\omega _0^z;s\right)+\frac{19}{9} J\left(\Omega _0,\omega _4^{z_1},\omega _{7,1}^{z_1},\omega _0^z;s\right) \\ -\frac{2}{3}
J\left(\Omega _0,\omega _7^s,\omega _0^y,\omega _0^y;s\right)-\frac{2}{3} J\left(\Omega _0,\omega _7^s,\omega _0^z,\omega _0^y;s\right)+\frac{2}{3} J\left(\Omega
_0,\omega _7^s,\omega _0^z,\omega _0^z;s\right)+\frac{2}{3} J\left(\Omega _0,\omega _7^s,\omega _y^z,\omega _0^y;s\right) \\ -\frac{2}{3} J\left(\Omega _0,\omega
_7^s,\omega _y^z,\omega _0^z;s\right)+\frac{2}{3} J\left(\Omega _0,\omega _7^s,\omega _{\bar{y}}^z,\omega _0^y;s\right)+\frac{2}{3} J\left(\Omega _0,\omega
_7^s,\omega _{\bar{y}}^z,\omega _0^z;s\right)-\frac{5}{12} J\left(\Omega _1,\omega _1^{z_1},\omega _{1,y}^{z_1},\omega _0^y;s\right) \\ +\frac{5}{12} J\left(\Omega
_1,\omega _1^{z_1},\omega _{1,y}^{z_1},\omega _0^z;s\right)+\frac{16}{9} J\left(\Omega _1,\omega _1^{z_1},\omega _{4,y}^{z_1},\omega _0^y;s\right)-\frac{16}{9}
J\left(\Omega _1,\omega _1^{z_1},\omega _{4,y}^{z_1},\omega _0^z;s\right)-\frac{19}{9} J\left(\Omega _1,\omega _1^{z_1},\omega _{7,y}^{z_1},\omega
_0^y;s\right) \\ +\frac{19}{9} J\left(\Omega _1,\omega _1^{z_1},\omega _{7,y}^{z_1},\omega _0^z;s\right)+\frac{3}{4} J\left(\Omega _1,\omega _1^{z_1},\omega
_{9,y}^{z_1},\omega _0^y;s\right)-\frac{3}{4} J\left(\Omega _1,\omega _1^{z_1},\omega _{9,y}^{z_1},\omega _0^z;s\right)+\frac{5}{12} J\left(\Omega _1,\omega
_4^{z_1},\omega _{1,y}^{z_1},\omega _0^y;s\right) \\ -\frac{5}{12} J\left(\Omega _1,\omega _4^{z_1},\omega _{1,y}^{z_1},\omega _0^z;s\right)-\frac{16}{9}
J\left(\Omega _1,\omega _4^{z_1},\omega _{4,y}^{z_1},\omega _0^y;s\right)+\frac{16}{9} J\left(\Omega _1,\omega _4^{z_1},\omega _{4,y}^{z_1},\omega
_0^z;s\right)+\frac{19}{9} J\left(\Omega _1,\omega _4^{z_1},\omega _{7,y}^{z_1},\omega _0^y;s\right) \\ -\frac{19}{9} J\left(\Omega _1,\omega _4^{z_1},\omega
_{7,y}^{z_1},\omega _0^z;s\right)-\frac{3}{4} J\left(\Omega _1,\omega _4^{z_1},\omega _{9,y}^{z_1},\omega _0^y;s\right)+\frac{3}{4} J\left(\Omega _1,\omega
_4^{z_1},\omega _{9,y}^{z_1},\omega _0^z;s\right)+\frac{5}{12} J\left(\Omega _2,\omega _1^{z_1},\omega _{1,\bar{y}}^{z_1},\omega _0^y;s\right) \\ +\frac{5}{12}
J\left(\Omega _2,\omega _1^{z_1},\omega _{1,\bar{y}}^{z_1},\omega _0^z;s\right)-\frac{4}{3} J\left(\Omega _2,\omega _1^{z_1},\omega _{4,\bar{y}}^{z_1},\omega
_0^y;s\right)-\frac{4}{3} J\left(\Omega _2,\omega _1^{z_1},\omega _{4,\bar{y}}^{z_1},\omega _0^z;s\right)-\frac{3}{4} J\left(\Omega _2,\omega _1^{z_1},\omega
_{9,\bar{y}}^{z_1},\omega _0^y;s\right) \\ -\frac{3}{4} J\left(\Omega _2,\omega _1^{z_1},\omega _{9,\bar{y}}^{z_1},\omega _0^z;s\right)-\frac{5}{12} J\left(\Omega
_2,\omega _4^{z_1},\omega _{1,\bar{y}}^{z_1},\omega _0^y;s\right)-\frac{5}{12} J\left(\Omega _2,\omega _4^{z_1},\omega _{1,\bar{y}}^{z_1},\omega
_0^z;s\right)+\frac{4}{3} J\left(\Omega _2,\omega _4^{z_1},\omega _{4,\bar{y}}^{z_1},\omega _0^y;s\right) \\ +\frac{4}{3} J\left(\Omega _2,\omega _4^{z_1},\omega
_{4,\bar{y}}^{z_1},\omega _0^z;s\right)+\frac{3}{4} J\left(\Omega _2,\omega _4^{z_1},\omega _{9,\bar{y}}^{z_1},\omega _0^y;s\right)+\frac{3}{4} J\left(\Omega
_2,\omega _4^{z_1},\omega _{9,\bar{y}}^{z_1},\omega _0^z;s\right)+\frac{4}{9} J\left(\Omega _3,\omega _1^{z_1},\omega _{4,1}^{z_1},\omega
_0^z;s\right) \\ +\frac{2}{9} J\left(\Omega _3,\omega _1^{z_1},\omega _{4,y}^{z_1},\omega _0^y;s\right)-\frac{2}{9} J\left(\Omega _3,\omega _1^{z_1},\omega
_{4,y}^{z_1},\omega _0^z;s\right)-\frac{2}{9} J\left(\Omega _3,\omega _1^{z_1},\omega _{4,\bar{y}}^{z_1},\omega _0^y;s\right)-\frac{2}{9} J\left(\Omega _3,\omega
_1^{z_1},\omega _{4,\bar{y}}^{z_1},\omega _0^z;s\right) \\ -\frac{19}{9} J\left(\Omega _3,\omega _1^{z_1},\omega _{7,1}^{z_1},\omega _0^z;s\right)-\frac{19}{18}
J\left(\Omega _3,\omega _1^{z_1},\omega _{7,y}^{z_1},\omega _0^y;s\right)+\frac{19}{18} J\left(\Omega _3,\omega _1^{z_1},\omega _{7,y}^{z_1},\omega
_0^z;s\right)  +\frac{19}{18} J\left(\Omega _3,\omega _1^{z_1},\omega _{7,\bar{y}}^{z_1},\omega _0^y;s\right) \\ +\frac{19}{18} J\left(\Omega _3,\omega _1^{z_1},\omega
_{7,\bar{y}}^{z_1},\omega _0^z;s\right) -\frac{4}{9} J\left(\Omega _3,\omega _4^{z_1},\omega _{4,1}^{z_1},\omega _0^z;s\right)-\frac{2}{9} J\left(\Omega _3,\omega
_4^{z_1},\omega _{4,y}^{z_1},\omega _0^y;s\right)+\frac{2}{9} J\left(\Omega _3,\omega _4^{z_1},\omega _{4,y}^{z_1},\omega _0^z;s\right) \\ +\frac{2}{9} J\left(\Omega
_3,\omega _4^{z_1},\omega _{4,\bar{y}}^{z_1},\omega _0^y;s\right)+\frac{2}{9} J\left(\Omega _3,\omega _4^{z_1},\omega _{4,\bar{y}}^{z_1},\omega
_0^z;s\right)+\frac{19}{9} J\left(\Omega _3,\omega _4^{z_1},\omega _{7,1}^{z_1},\omega _0^z;s\right)+\frac{19}{18} J\left(\Omega _3,\omega _4^{z_1},\omega
_{7,y}^{z_1},\omega _0^y;s\right) \\ -\frac{19}{18} J\left(\Omega _3,\omega _4^{z_1},\omega _{7,y}^{z_1},\omega _0^z;s\right)-\frac{19}{18} J\left(\Omega _3,\omega
_4^{z_1},\omega _{7,\bar{y}}^{z_1},\omega _0^y;s\right)-\frac{19}{18} J\left(\Omega _3,\omega _4^{z_1},\omega _{7,\bar{y}}^{z_1},\omega _0^z;s\right)+\frac{4}{9}
J\left(\Omega _4,\omega _1^{z_1},\omega _{4,-1}^{z_1},\omega _0^z;s\right) \\ +\frac{2}{9} J\left(\Omega _4,\omega _1^{z_1},\omega _{4,y}^{z_1},\omega
_0^y;s\right)-\frac{2}{9} J\left(\Omega _4,\omega _1^{z_1},\omega _{4,y}^{z_1},\omega _0^z;s\right)-\frac{2}{9} J\left(\Omega _4,\omega _1^{z_1},\omega
_{4,\bar{y}}^{z_1},\omega _0^y;s\right)-\frac{2}{9} J\left(\Omega _4,\omega _1^{z_1},\omega _{4,\bar{y}}^{z_1},\omega _0^z;s\right) \\ -\frac{19}{9} J\left(\Omega
_4,\omega _1^{z_1},\omega _{7,-1}^{z_1},\omega _0^z;s\right)-\frac{19}{18} J\left(\Omega _4,\omega _1^{z_1},\omega _{7,y}^{z_1},\omega
_0^y;s\right)+\frac{19}{18} J\left(\Omega _4,\omega _1^{z_1},\omega _{7,y}^{z_1},\omega _0^z;s\right)  +\frac{19}{18} J\left(\Omega _4,\omega _1^{z_1},\omega
_{7,\bar{y}}^{z_1},\omega _0^y;s\right) \\ +\frac{19}{18} J\left(\Omega _4,\omega _1^{z_1},\omega _{7,\bar{y}}^{z_1},\omega _0^z;s\right)-\frac{4}{9} J\left(\Omega
_4,\omega _4^{z_1},\omega _{4,-1}^{z_1},\omega _0^z;s\right)-\frac{2}{9} J\left(\Omega _4,\omega _4^{z_1},\omega _{4,y}^{z_1},\omega _0^y;s\right)+\frac{2}{9}
J\left(\Omega _4,\omega _4^{z_1},\omega _{4,y}^{z_1},\omega _0^z;s\right) \\ +\frac{2}{9} J\left(\Omega _4,\omega _4^{z_1},\omega _{4,\bar{y}}^{z_1},\omega
_0^y;s\right)+\frac{2}{9} J\left(\Omega _4,\omega _4^{z_1},\omega _{4,\bar{y}}^{z_1},\omega _0^z;s\right)+\frac{19}{9} J\left(\Omega _4,\omega _4^{z_1},\omega
_{7,-1}^{z_1},\omega _0^z;s\right)+\frac{19}{18} J\left(\Omega _4,\omega _4^{z_1},\omega _{7,y}^{z_1},\omega _0^y;s\right) \\ -\frac{19}{18} J\left(\Omega _4,\omega
_4^{z_1},\omega _{7,y}^{z_1},\omega _0^z;s\right)-\frac{19}{18} J\left(\Omega _4,\omega _4^{z_1},\omega _{7,\bar{y}}^{z_1},\omega _0^y;s\right)-\frac{19}{18}
J\left(\Omega _4,\omega _4^{z_1},\omega _{7,\bar{y}}^{z_1},\omega _0^z;s\right)+\frac{5}{12} J\left(\Omega _0,\omega _1^{z_1},\omega _1^{z_1},\omega _0^y,\omega
_0^y;s\right) \\ +\frac{5}{12} J\left(\Omega _0,\omega _1^{z_1},\omega _1^{z_1},\omega _0^z,\omega _0^y;s\right)-\frac{5}{12} J\left(\Omega _0,\omega _1^{z_1},\omega
_1^{z_1},\omega _0^z,\omega _0^z;s\right)-\frac{5}{12} J\left(\Omega _0,\omega _1^{z_1},\omega _1^{z_1},\omega _y^z,\omega _0^y;s\right) \\ +\frac{5}{12}
J\left(\Omega _0,\omega _1^{z_1},\omega _1^{z_1},\omega _y^z,\omega _0^z;s\right)-\frac{5}{12} J\left(\Omega _0,\omega _1^{z_1},\omega _1^{z_1},\omega
_{\bar{y}}^z,\omega _0^y;s\right)-\frac{5}{12} J\left(\Omega _0,\omega _1^{z_1},\omega _1^{z_1},\omega _{\bar{y}}^z,\omega _0^z;s\right) \\ -\frac{16}{9}
J\left(\Omega _0,\omega _1^{z_1},\omega _4^{z_1},\omega _0^y,\omega _0^y;s\right)-\frac{16}{9} J\left(\Omega _0,\omega _1^{z_1},\omega _4^{z_1},\omega
_0^z,\omega _0^y;s\right) +\frac{16}{9} J\left(\Omega _0,\omega _1^{z_1},\omega _4^{z_1},\omega _0^z,\omega _0^z;s\right) \\ +\frac{16}{9} J\left(\Omega _0,\omega
_1^{z_1},\omega _4^{z_1},\omega _y^z,\omega _0^y;s\right)-\frac{16}{9} J\left(\Omega _0,\omega _1^{z_1},\omega _4^{z_1},\omega _y^z,\omega
_0^z;s\right)+\frac{16}{9} J\left(\Omega _0,\omega _1^{z_1},\omega _4^{z_1},\omega _{\bar{y}}^z,\omega _0^y;s\right) \\ +\frac{16}{9} J\left(\Omega _0,\omega
_1^{z_1},\omega _4^{z_1},\omega _{\bar{y}}^z,\omega _0^z;s\right)+\frac{19}{9} J\left(\Omega _0,\omega _1^{z_1},\omega _7^{z_1},\omega _0^y,\omega
_0^y;s\right)+\frac{19}{9} J\left(\Omega _0,\omega _1^{z_1},\omega _7^{z_1},\omega _0^z,\omega _0^y;s\right) \\ -\frac{19}{9} J\left(\Omega _0,\omega _1^{z_1},\omega
_7^{z_1},\omega _0^z,\omega _0^z;s\right)-\frac{19}{9} J\left(\Omega _0,\omega _1^{z_1},\omega _7^{z_1},\omega _y^z,\omega _0^y;s\right)+\frac{19}{9}
J\left(\Omega _0,\omega _1^{z_1},\omega _7^{z_1},\omega _y^z,\omega _0^z;s\right) \\ -\frac{19}{9} J\left(\Omega _0,\omega _1^{z_1},\omega _7^{z_1},\omega
_{\bar{y}}^z,\omega _0^y;s\right)-\frac{19}{9} J\left(\Omega _0,\omega _1^{z_1},\omega _7^{z_1},\omega _{\bar{y}}^z,\omega _0^z;s\right)  -\frac{3}{4}
J\left(\Omega _0,\omega _1^{z_1},\omega _9^{z_1},\omega _0^y,\omega _0^y;s\right) \\ -\frac{3}{4} J\left(\Omega _0,\omega _1^{z_1},\omega _9^{z_1},\omega _0^z,\omega
_0^y;s\right)+\frac{3}{4} J\left(\Omega _0,\omega _1^{z_1},\omega _9^{z_1},\omega _0^z,\omega _0^z;s\right)+\frac{3}{4} J\left(\Omega _0,\omega _1^{z_1},\omega
_9^{z_1},\omega _y^z,\omega _0^y;s\right) \\ -\frac{3}{4} J\left(\Omega _0,\omega _1^{z_1},\omega _9^{z_1},\omega _y^z,\omega _0^z;s\right)+\frac{3}{4} J\left(\Omega
_0,\omega _1^{z_1},\omega _9^{z_1},\omega _{\bar{y}}^z,\omega _0^y;s\right)+\frac{3}{4} J\left(\Omega _0,\omega _1^{z_1},\omega _9^{z_1},\omega
_{\bar{y}}^z,\omega _0^z;s\right)\\ -\frac{5}{12} J\left(\Omega _0,\omega _4^{z_1},\omega _1^{z_1},\omega _0^y,\omega _0^y;s\right)-\frac{5}{12} J\left(\Omega
_0,\omega _4^{z_1},\omega _1^{z_1},\omega _0^z,\omega _0^y;s\right)+\frac{5}{12} J\left(\Omega _0,\omega _4^{z_1},\omega _1^{z_1},\omega _0^z,\omega
_0^z;s\right)\\ +\frac{5}{12} J\left(\Omega _0,\omega _4^{z_1},\omega _1^{z_1},\omega _y^z,\omega _0^y;s\right)-\frac{5}{12} J\left(\Omega _0,\omega _4^{z_1},\omega
_1^{z_1},\omega _y^z,\omega _0^z;s\right)+\frac{5}{12} J\left(\Omega _0,\omega _4^{z_1},\omega _1^{z_1},\omega _{\bar{y}}^z,\omega _0^y;s\right) \\ +\frac{5}{12}
J\left(\Omega _0,\omega _4^{z_1},\omega _1^{z_1},\omega _{\bar{y}}^z,\omega _0^z;s\right)+\frac{16}{9} J\left(\Omega _0,\omega _4^{z_1},\omega _4^{z_1},\omega
_0^y,\omega _0^y;s\right)+\frac{16}{9} J\left(\Omega _0,\omega _4^{z_1},\omega _4^{z_1},\omega _0^z,\omega _0^y;s\right) \\ -\frac{16}{9} J\left(\Omega _0,\omega
_4^{z_1},\omega _4^{z_1},\omega _0^z,\omega _0^z;s\right)-\frac{16}{9} J\left(\Omega _0,\omega _4^{z_1},\omega _4^{z_1},\omega _y^z,\omega
_0^y;s\right)+\frac{16}{9} J\left(\Omega _0,\omega _4^{z_1},\omega _4^{z_1},\omega _y^z,\omega _0^z;s\right) \\ -\frac{16}{9} J\left(\Omega _0,\omega _4^{z_1},\omega
_4^{z_1},\omega _{\bar{y}}^z,\omega _0^y;s\right)-\frac{16}{9} J\left(\Omega _0,\omega _4^{z_1},\omega _4^{z_1},\omega _{\bar{y}}^z,\omega
_0^z;s\right)  -\frac{19}{9} J\left(\Omega _0,\omega _4^{z_1},\omega _7^{z_1},\omega _0^y,\omega _0^y;s\right) \\ -\frac{19}{9} J\left(\Omega _0,\omega _4^{z_1},\omega
_7^{z_1},\omega _0^z,\omega _0^y;s\right)+\frac{19}{9} J\left(\Omega _0,\omega _4^{z_1},\omega _7^{z_1},\omega _0^z,\omega _0^z;s\right)+\frac{19}{9}
J\left(\Omega _0,\omega _4^{z_1},\omega _7^{z_1},\omega _y^z,\omega _0^y;s\right) \\ -\frac{19}{9} J\left(\Omega _0,\omega _4^{z_1},\omega _7^{z_1},\omega
_y^z,\omega _0^z;s\right)+\frac{19}{9} J\left(\Omega _0,\omega _4^{z_1},\omega _7^{z_1},\omega _{\bar{y}}^z,\omega _0^y;s\right)+\frac{19}{9} J\left(\Omega
_0,\omega _4^{z_1},\omega _7^{z_1},\omega _{\bar{y}}^z,\omega _0^z;s\right) \\ +\frac{3}{4} J\left(\Omega _0,\omega _4^{z_1},\omega _9^{z_1},\omega _0^y,\omega
_0^y;s\right)+\frac{3}{4} J\left(\Omega _0,\omega _4^{z_1},\omega _9^{z_1},\omega _0^z,\omega _0^y;s\right)-\frac{3}{4} J\left(\Omega _0,\omega _4^{z_1},\omega
_9^{z_1},\omega _0^z,\omega _0^z;s\right) \\ -\frac{3}{4} J\left(\Omega _0,\omega _4^{z_1},\omega _9^{z_1},\omega _y^z,\omega _0^y;s\right)+\frac{3}{4} J\left(\Omega
_0,\omega _4^{z_1},\omega _9^{z_1},\omega _y^z,\omega _0^z;s\right)-\frac{3}{4} J\left(\Omega _0,\omega _4^{z_1},\omega _9^{z_1},\omega _{\bar{y}}^z,\omega
_0^y;s\right) \\ -\frac{3}{4} J\left(\Omega _0,\omega _4^{z_1},\omega _9^{z_1},\omega _{\bar{y}}^z,\omega _0^z;s\right)\Big] + \OO(\ep).
\end{multline}

\begin{multline}
e^{2\ep\gamma_E}j_3^{\kite}(1,1,1,1,1) =\frac{1}{s}\Big[\left(-8+\frac{\pi ^2}{3}\right) G(3,s)+\left(4-\frac{\pi ^2}{6}\right) G(7,s)+\frac{5}{24} \left(-12+\pi ^2\right) J\left(\Omega _0,\omega _3^{z_1},\omega
_1^{z_1};s\right) \\ -\frac{8}{9} \left(-15+\pi ^2\right) J\left(\Omega _0,\omega _3^{z_1},\omega _4^{z_1};s\right)+\frac{19}{18} \left(-24+\pi ^2\right)
J\left(\Omega _0,\omega _3^{z_1},\omega _7^{z_1};s\right)-\frac{3}{8} \left(-12+\pi ^2\right) J\left(\Omega _0,\omega _3^{z_1},\omega
_9^{z_1};s\right) \\ -\frac{5}{24} \left(-12+\pi ^2\right) J\left(\Omega _0,\omega _4^{z_1},\omega _1^{z_1};s\right)+\frac{8}{9} \left(-15+\pi ^2\right)
J\left(\Omega _0,\omega _4^{z_1},\omega _4^{z_1};s\right)-\frac{19}{18} \left(-24+\pi ^2\right) J\left(\Omega _0,\omega _4^{z_1},\omega
_7^{z_1};s\right) \\ +\frac{3}{8} \left(-12+\pi ^2\right) J\left(\Omega _0,\omega _4^{z_1},\omega _9^{z_1};s\right)+4 J\left(\Omega _0,\omega _{3,-1}^s,\omega
_0^z;s\right)-4 J\left(\Omega _0,\omega _{3,1}^s,\omega _0^z;s\right)-2 J\left(\Omega _0,\omega _{7,-1}^s,\omega _0^z;s\right) \\ +2 J\left(\Omega _0,\omega
_{7,1}^s,\omega _0^z;s\right)-4 J\left(\Omega _1,\omega _{3,y}^s,\omega _0^y;s\right)+4 J\left(\Omega _1,\omega _{3,y}^s,\omega _0^z;s\right)+2 J\left(\Omega
_1,\omega _{7,y}^s,\omega _0^y;s\right)-2 J\left(\Omega _1,\omega _{7,y}^s,\omega _0^z;s\right) \\ -4 J\left(\Omega _3,\omega _{3,1}^s,\omega _0^z;s\right)-2
J\left(\Omega _3,\omega _{3,y}^s,\omega _0^y;s\right)+2 J\left(\Omega _3,\omega _{3,y}^s,\omega _0^z;s\right)+2 J\left(\Omega _3,\omega _{3,\bar{y}}^s,\omega
_0^y;s\right)+2 J\left(\Omega _3,\omega _{3,\bar{y}}^s,\omega _0^z;s\right) \\ +2 J\left(\Omega _3,\omega _{7,1}^s,\omega _0^z;s\right)+J\left(\Omega _3,\omega
_{7,y}^s,\omega _0^y;s\right)-J\left(\Omega _3,\omega _{7,y}^s,\omega _0^z;s\right)-J\left(\Omega _3,\omega _{7,\bar{y}}^s,\omega _0^y;s\right)-J\left(\Omega
_3,\omega _{7,\bar{y}}^s,\omega _0^z;s\right) \\ -4 J\left(\Omega _4,\omega _{3,-1}^s,\omega _0^z;s\right)-2 J\left(\Omega _4,\omega _{3,y}^s,\omega _0^y;s\right)+2
J\left(\Omega _4,\omega _{3,y}^s,\omega _0^z;s\right)+2 J\left(\Omega _4,\omega _{3,\bar{y}}^s,\omega _0^y;s\right)+2 J\left(\Omega _4,\omega
_{3,\bar{y}}^s,\omega _0^z;s\right) \\ +2 J\left(\Omega _4,\omega _{7,-1}^s,\omega _0^z;s\right)+J\left(\Omega _4,\omega _{7,y}^s,\omega _0^y;s\right)-J\left(\Omega
_4,\omega _{7,y}^s,\omega _0^z;s\right)-J\left(\Omega _4,\omega _{7,\bar{y}}^s,\omega _0^y;s\right)-J\left(\Omega _4,\omega _{7,\bar{y}}^s,\omega _0^z;s\right) \\ +4
J\left(\Omega _0,\omega _3^s,\omega _0^y,\omega _0^y;s\right) +4 J\left(\Omega _0,\omega _3^s,\omega _0^z,\omega _0^y;s\right)-4 J\left(\Omega _0,\omega
_3^s,\omega _0^z,\omega _0^z;s\right)-2 J\left(\Omega _0,\omega _3^s,\omega _4^{z_1},\omega _4^{z_1};s\right) \\ -4 J\left(\Omega _0,\omega _3^s,\omega _y^z,\omega
_0^y;s\right)+4 J\left(\Omega _0,\omega _3^s,\omega _y^z,\omega _0^z;s\right)-4 J\left(\Omega _0,\omega _3^s,\omega _{\bar{y}}^z,\omega _0^y;s\right)-4
J\left(\Omega _0,\omega _3^s,\omega _{\bar{y}}^z,\omega _0^z;s\right) \\ +\frac{5}{2} J\left(\Omega _0,\omega _3^{z_1},\omega _1^{z_1},\omega _0^z;s\right)-8
J\left(\Omega _0,\omega _3^{z_1},\omega _4^{z_1},\omega _0^z;s\right)-\frac{9}{2} J\left(\Omega _0,\omega _3^{z_1},\omega _9^{z_1},\omega
_0^z;s\right)-\frac{8}{3} J\left(\Omega _0,\omega _3^{z_1},\omega _{4,-1}^{z_1},\omega _0^z;s\right) \\ +\frac{8}{3} J\left(\Omega _0,\omega _3^{z_1},\omega
_{4,1}^{z_1},\omega _0^z;s\right)+\frac{38}{3} J\left(\Omega _0,\omega _3^{z_1},\omega _{7,-1}^{z_1},\omega _0^z;s\right)-\frac{38}{3} J\left(\Omega _0,\omega
_3^{z_1},\omega _{7,1}^{z_1},\omega _0^z;s\right)-\frac{5}{2} J\left(\Omega _0,\omega _4^{z_1},\omega _1^{z_1},\omega _0^z;s\right) \\ +8 J\left(\Omega _0,\omega
_4^{z_1},\omega _4^{z_1},\omega _0^z;s\right)+\frac{9}{2} J\left(\Omega _0,\omega _4^{z_1},\omega _9^{z_1},\omega _0^z;s\right)+\frac{8}{3} J\left(\Omega
_0,\omega _4^{z_1},\omega _{4,-1}^{z_1},\omega _0^z;s\right)-\frac{8}{3} J\left(\Omega _0,\omega _4^{z_1},\omega _{4,1}^{z_1},\omega _0^z;s\right) \\ -\frac{38}{3}
J\left(\Omega _0,\omega _4^{z_1},\omega _{7,-1}^{z_1},\omega _0^z;s\right)+\frac{38}{3} J\left(\Omega _0,\omega _4^{z_1},\omega _{7,1}^{z_1},\omega
_0^z;s\right)-2 J\left(\Omega _0,\omega _7^s,\omega _0^y,\omega _0^y;s\right)-2 J\left(\Omega _0,\omega _7^s,\omega _0^z,\omega _0^y;s\right) \\ +2 J\left(\Omega
_0,\omega _7^s,\omega _0^z,\omega _0^z;s\right)+2 J\left(\Omega _0,\omega _7^s,\omega _y^z,\omega _0^y;s\right)-2 J\left(\Omega _0,\omega _7^s,\omega _y^z,\omega
_0^z;s\right)+2 J\left(\Omega _0,\omega _7^s,\omega _{\bar{y}}^z,\omega _0^y;s\right) \\ +2 J\left(\Omega _0,\omega _7^s,\omega _{\bar{y}}^z,\omega
_0^z;s\right)-\frac{5}{2} J\left(\Omega _1,\omega _3^{z_1},\omega _{1,y}^{z_1},\omega _0^y;s\right)+\frac{5}{2} J\left(\Omega _1,\omega _3^{z_1},\omega
_{1,y}^{z_1},\omega _0^z;s\right)+\frac{32}{3} J\left(\Omega _1,\omega _3^{z_1},\omega _{4,y}^{z_1},\omega _0^y;s\right) \\ -\frac{32}{3} J\left(\Omega _1,\omega
_3^{z_1},\omega _{4,y}^{z_1},\omega _0^z;s\right)-\frac{38}{3} J\left(\Omega _1,\omega _3^{z_1},\omega _{7,y}^{z_1},\omega _0^y;s\right)+\frac{38}{3}
J\left(\Omega _1,\omega _3^{z_1},\omega _{7,y}^{z_1},\omega _0^z;s\right)+\frac{9}{2} J\left(\Omega _1,\omega _3^{z_1},\omega _{9,y}^{z_1},\omega
_0^y;s\right) \\ -\frac{9}{2} J\left(\Omega _1,\omega _3^{z_1},\omega _{9,y}^{z_1},\omega _0^z;s\right)+\frac{5}{2} J\left(\Omega _1,\omega _4^{z_1},\omega
_{1,y}^{z_1},\omega _0^y;s\right)-\frac{5}{2} J\left(\Omega _1,\omega _4^{z_1},\omega _{1,y}^{z_1},\omega _0^z;s\right)-\frac{32}{3} J\left(\Omega _1,\omega
_4^{z_1},\omega _{4,y}^{z_1},\omega _0^y;s\right) \\ +\frac{32}{3} J\left(\Omega _1,\omega _4^{z_1},\omega _{4,y}^{z_1},\omega _0^z;s\right)+\frac{38}{3}
J\left(\Omega _1,\omega _4^{z_1},\omega _{7,y}^{z_1},\omega _0^y;s\right)-\frac{38}{3} J\left(\Omega _1,\omega _4^{z_1},\omega _{7,y}^{z_1},\omega
_0^z;s\right)-\frac{9}{2} J\left(\Omega _1,\omega _4^{z_1},\omega _{9,y}^{z_1},\omega _0^y;s\right) \\ +\frac{9}{2} J\left(\Omega _1,\omega _4^{z_1},\omega
_{9,y}^{z_1},\omega _0^z;s\right)+\frac{5}{2} J\left(\Omega _2,\omega _3^{z_1},\omega _{1,\bar{y}}^{z_1},\omega _0^y;s\right)+\frac{5}{2} J\left(\Omega _2,\omega
_3^{z_1},\omega _{1,\bar{y}}^{z_1},\omega _0^z;s\right)-8 J\left(\Omega _2,\omega _3^{z_1},\omega _{4,\bar{y}}^{z_1},\omega _0^y;s\right) \\ -8 J\left(\Omega
_2,\omega _3^{z_1},\omega _{4,\bar{y}}^{z_1},\omega _0^z;s\right)-\frac{9}{2} J\left(\Omega _2,\omega _3^{z_1},\omega _{9,\bar{y}}^{z_1},\omega
_0^y;s\right)-\frac{9}{2} J\left(\Omega _2,\omega _3^{z_1},\omega _{9,\bar{y}}^{z_1},\omega _0^z;s\right)-\frac{5}{2} J\left(\Omega _2,\omega _4^{z_1},\omega
_{1,\bar{y}}^{z_1},\omega _0^y;s\right) \\ -\frac{5}{2} J\left(\Omega _2,\omega _4^{z_1},\omega _{1,\bar{y}}^{z_1},\omega _0^z;s\right)+8 J\left(\Omega _2,\omega
_4^{z_1},\omega _{4,\bar{y}}^{z_1},\omega _0^y;s\right)+8 J\left(\Omega _2,\omega _4^{z_1},\omega _{4,\bar{y}}^{z_1},\omega _0^z;s\right)+\frac{9}{2}
J\left(\Omega _2,\omega _4^{z_1},\omega _{9,\bar{y}}^{z_1},\omega _0^y;s\right) \\ +\frac{9}{2} J\left(\Omega _2,\omega _4^{z_1},\omega _{9,\bar{y}}^{z_1},\omega
_0^z;s\right)+\frac{8}{3} J\left(\Omega _3,\omega _3^{z_1},\omega _{4,1}^{z_1},\omega _0^z;s\right)+\frac{4}{3} J\left(\Omega _3,\omega _3^{z_1},\omega
_{4,y}^{z_1},\omega _0^y;s\right)-\frac{4}{3} J\left(\Omega _3,\omega _3^{z_1},\omega _{4,y}^{z_1},\omega _0^z;s\right) \\ -\frac{4}{3} J\left(\Omega _3,\omega
_3^{z_1},\omega _{4,\bar{y}}^{z_1},\omega _0^y;s\right)-\frac{4}{3} J\left(\Omega _3,\omega _3^{z_1},\omega _{4,\bar{y}}^{z_1},\omega _0^z;s\right)-\frac{38}{3}
J\left(\Omega _3,\omega _3^{z_1},\omega _{7,1}^{z_1},\omega _0^z;s\right)-\frac{19}{3} J\left(\Omega _3,\omega _3^{z_1},\omega _{7,y}^{z_1},\omega
_0^y;s\right) \\ +\frac{19}{3} J\left(\Omega _3,\omega _3^{z_1},\omega _{7,y}^{z_1},\omega _0^z;s\right)+\frac{19}{3} J\left(\Omega _3,\omega _3^{z_1},\omega
_{7,\bar{y}}^{z_1},\omega _0^y;s\right)+\frac{19}{3} J\left(\Omega _3,\omega _3^{z_1},\omega _{7,\bar{y}}^{z_1},\omega _0^z;s\right)-\frac{8}{3} J\left(\Omega
_3,\omega _4^{z_1},\omega _{4,1}^{z_1},\omega _0^z;s\right) \\ -\frac{4}{3} J\left(\Omega _3,\omega _4^{z_1},\omega _{4,y}^{z_1},\omega _0^y;s\right)+\frac{4}{3}
J\left(\Omega _3,\omega _4^{z_1},\omega _{4,y}^{z_1},\omega _0^z;s\right)+\frac{4}{3} J\left(\Omega _3,\omega _4^{z_1},\omega _{4,\bar{y}}^{z_1},\omega
_0^y;s\right)+\frac{4}{3} J\left(\Omega _3,\omega _4^{z_1},\omega _{4,\bar{y}}^{z_1},\omega _0^z;s\right) \\ +\frac{38}{3} J\left(\Omega _3,\omega _4^{z_1},\omega
_{7,1}^{z_1},\omega _0^z;s\right)+\frac{19}{3} J\left(\Omega _3,\omega _4^{z_1},\omega _{7,y}^{z_1},\omega _0^y;s\right)-\frac{19}{3} J\left(\Omega _3,\omega
_4^{z_1},\omega _{7,y}^{z_1},\omega _0^z;s\right)-\frac{19}{3} J\left(\Omega _3,\omega _4^{z_1},\omega _{7,\bar{y}}^{z_1},\omega _0^y;s\right) \\ -\frac{19}{3}
J\left(\Omega _3,\omega _4^{z_1},\omega _{7,\bar{y}}^{z_1},\omega _0^z;s\right)+\frac{8}{3} J\left(\Omega _4,\omega _3^{z_1},\omega _{4,-1}^{z_1},\omega
_0^z;s\right)+\frac{4}{3} J\left(\Omega _4,\omega _3^{z_1},\omega _{4,y}^{z_1},\omega _0^y;s\right)-\frac{4}{3} J\left(\Omega _4,\omega _3^{z_1},\omega
_{4,y}^{z_1},\omega _0^z;s\right) \\ -\frac{4}{3} J\left(\Omega _4,\omega _3^{z_1},\omega _{4,\bar{y}}^{z_1},\omega _0^y;s\right)-\frac{4}{3} J\left(\Omega _4,\omega
_3^{z_1},\omega _{4,\bar{y}}^{z_1},\omega _0^z;s\right)-\frac{38}{3} J\left(\Omega _4,\omega _3^{z_1},\omega _{7,-1}^{z_1},\omega _0^z;s\right)-\frac{19}{3}
J\left(\Omega _4,\omega _3^{z_1},\omega _{7,y}^{z_1},\omega _0^y;s\right) \\ +\frac{19}{3} J\left(\Omega _4,\omega _3^{z_1},\omega _{7,y}^{z_1},\omega
_0^z;s\right)+\frac{19}{3} J\left(\Omega _4,\omega _3^{z_1},\omega _{7,\bar{y}}^{z_1},\omega _0^y;s\right)+\frac{19}{3} J\left(\Omega _4,\omega _3^{z_1},\omega
_{7,\bar{y}}^{z_1},\omega _0^z;s\right)-\frac{8}{3} J\left(\Omega _4,\omega _4^{z_1},\omega _{4,-1}^{z_1},\omega _0^z;s\right) \\ -\frac{4}{3} J\left(\Omega
_4,\omega _4^{z_1},\omega _{4,y}^{z_1},\omega _0^y;s\right)+\frac{4}{3} J\left(\Omega _4,\omega _4^{z_1},\omega _{4,y}^{z_1},\omega _0^z;s\right)+\frac{4}{3}
J\left(\Omega _4,\omega _4^{z_1},\omega _{4,\bar{y}}^{z_1},\omega _0^y;s\right)+\frac{4}{3} J\left(\Omega _4,\omega _4^{z_1},\omega _{4,\bar{y}}^{z_1},\omega
_0^z;s\right) \\ +\frac{38}{3} J\left(\Omega _4,\omega _4^{z_1},\omega _{7,-1}^{z_1},\omega _0^z;s\right)+\frac{19}{3} J\left(\Omega _4,\omega _4^{z_1},\omega
_{7,y}^{z_1},\omega _0^y;s\right)-\frac{19}{3} J\left(\Omega _4,\omega _4^{z_1},\omega _{7,y}^{z_1},\omega _0^z;s\right)-\frac{19}{3} J\left(\Omega _4,\omega
_4^{z_1},\omega _{7,\bar{y}}^{z_1},\omega _0^y;s\right) \\ -\frac{19}{3} J\left(\Omega _4,\omega _4^{z_1},\omega _{7,\bar{y}}^{z_1},\omega _0^z;s\right)+\frac{5}{2}
J\left(\Omega _0,\omega _3^{z_1},\omega _1^{z_1},\omega _0^y,\omega _0^y;s\right)+\frac{5}{2} J\left(\Omega _0,\omega _3^{z_1},\omega _1^{z_1},\omega _0^z,\omega
_0^y;s\right)-\frac{5}{2} J\left(\Omega _0,\omega _3^{z_1},\omega _1^{z_1},\omega _0^z,\omega _0^z;s\right) \\ -\frac{5}{2} J\left(\Omega _0,\omega _3^{z_1},\omega
_1^{z_1},\omega _y^z,\omega _0^y;s\right)+\frac{5}{2} J\left(\Omega _0,\omega _3^{z_1},\omega _1^{z_1},\omega _y^z,\omega _0^z;s\right)-\frac{5}{2} J\left(\Omega
_0,\omega _3^{z_1},\omega _1^{z_1},\omega _{\bar{y}}^z,\omega _0^y;s\right) \\ -\frac{5}{2} J\left(\Omega _0,\omega _3^{z_1},\omega _1^{z_1},\omega
_{\bar{y}}^z,\omega _0^z;s\right)-\frac{32}{3} J\left(\Omega _0,\omega _3^{z_1},\omega _4^{z_1},\omega _0^y,\omega _0^y;s\right)-\frac{32}{3} J\left(\Omega
_0,\omega _3^{z_1},\omega _4^{z_1},\omega _0^z,\omega _0^y;s\right) \\ +\frac{32}{3} J\left(\Omega _0,\omega _3^{z_1},\omega _4^{z_1},\omega _0^z,\omega
_0^z;s\right)+\frac{32}{3} J\left(\Omega _0,\omega _3^{z_1},\omega _4^{z_1},\omega _y^z,\omega _0^y;s\right)-\frac{32}{3} J\left(\Omega _0,\omega _3^{z_1},\omega
_4^{z_1},\omega _y^z,\omega _0^z;s\right) \\ +\frac{32}{3} J\left(\Omega _0,\omega _3^{z_1},\omega _4^{z_1},\omega _{\bar{y}}^z,\omega _0^y;s\right)+\frac{32}{3}
J\left(\Omega _0,\omega _3^{z_1},\omega _4^{z_1},\omega _{\bar{y}}^z,\omega _0^z;s\right)+\frac{38}{3} J\left(\Omega _0,\omega _3^{z_1},\omega _7^{z_1},\omega
_0^y,\omega _0^y;s\right) \\ +\frac{38}{3} J\left(\Omega _0,\omega _3^{z_1},\omega _7^{z_1},\omega _0^z,\omega _0^y;s\right)-\frac{38}{3} J\left(\Omega _0,\omega
_3^{z_1},\omega _7^{z_1},\omega _0^z,\omega _0^z;s\right)-\frac{38}{3} J\left(\Omega _0,\omega _3^{z_1},\omega _7^{z_1},\omega _y^z,\omega
_0^y;s\right) \\ +\frac{38}{3} J\left(\Omega _0,\omega _3^{z_1},\omega _7^{z_1},\omega _y^z,\omega _0^z;s\right)-\frac{38}{3} J\left(\Omega _0,\omega _3^{z_1},\omega
_7^{z_1},\omega _{\bar{y}}^z,\omega _0^y;s\right)-\frac{38}{3} J\left(\Omega _0,\omega _3^{z_1},\omega _7^{z_1},\omega _{\bar{y}}^z,\omega
_0^z;s\right) \\ -\frac{9}{2} J\left(\Omega _0,\omega _3^{z_1},\omega _9^{z_1},\omega _0^y,\omega _0^y;s\right)-\frac{9}{2} J\left(\Omega _0,\omega _3^{z_1},\omega
_9^{z_1},\omega _0^z,\omega _0^y;s\right)+\frac{9}{2} J\left(\Omega _0,\omega _3^{z_1},\omega _9^{z_1},\omega _0^z,\omega _0^z;s\right) \\ +\frac{9}{2} J\left(\Omega
_0,\omega _3^{z_1},\omega _9^{z_1},\omega _y^z,\omega _0^y;s\right)-\frac{9}{2} J\left(\Omega _0,\omega _3^{z_1},\omega _9^{z_1},\omega _y^z,\omega
_0^z;s\right)+\frac{9}{2} J\left(\Omega _0,\omega _3^{z_1},\omega _9^{z_1},\omega _{\bar{y}}^z,\omega _0^y;s\right) \\ +\frac{9}{2} J\left(\Omega _0,\omega
_3^{z_1},\omega _9^{z_1},\omega _{\bar{y}}^z,\omega _0^z;s\right)-\frac{5}{2} J\left(\Omega _0,\omega _4^{z_1},\omega _1^{z_1},\omega _0^y,\omega
_0^y;s\right)-\frac{5}{2} J\left(\Omega _0,\omega _4^{z_1},\omega _1^{z_1},\omega _0^z,\omega _0^y;s\right) \\ +\frac{5}{2} J\left(\Omega _0,\omega _4^{z_1},\omega
_1^{z_1},\omega _0^z,\omega _0^z;s\right)+\frac{5}{2} J\left(\Omega _0,\omega _4^{z_1},\omega _1^{z_1},\omega _y^z,\omega _0^y;s\right)-\frac{5}{2} J\left(\Omega
_0,\omega _4^{z_1},\omega _1^{z_1},\omega _y^z,\omega _0^z;s\right) \\ +\frac{5}{2} J\left(\Omega _0,\omega _4^{z_1},\omega _1^{z_1},\omega _{\bar{y}}^z,\omega
_0^y;s\right)+\frac{5}{2} J\left(\Omega _0,\omega _4^{z_1},\omega _1^{z_1},\omega _{\bar{y}}^z,\omega _0^z;s\right)+\frac{32}{3} J\left(\Omega _0,\omega
_4^{z_1},\omega _4^{z_1},\omega _0^y,\omega _0^y;s\right) \\ +\frac{32}{3} J\left(\Omega _0,\omega _4^{z_1},\omega _4^{z_1},\omega _0^z,\omega
_0^y;s\right)-\frac{32}{3} J\left(\Omega _0,\omega _4^{z_1},\omega _4^{z_1},\omega _0^z,\omega _0^z;s\right)-\frac{32}{3} J\left(\Omega _0,\omega _4^{z_1},\omega
_4^{z_1},\omega _y^z,\omega _0^y;s\right) \\ +\frac{32}{3} J\left(\Omega _0,\omega _4^{z_1},\omega _4^{z_1},\omega _y^z,\omega _0^z;s\right)-\frac{32}{3}
J\left(\Omega _0,\omega _4^{z_1},\omega _4^{z_1},\omega _{\bar{y}}^z,\omega _0^y;s\right)-\frac{32}{3} J\left(\Omega _0,\omega _4^{z_1},\omega _4^{z_1},\omega
_{\bar{y}}^z,\omega _0^z;s\right) \\ -\frac{38}{3} J\left(\Omega _0,\omega _4^{z_1},\omega _7^{z_1},\omega _0^y,\omega _0^y;s\right)-\frac{38}{3} J\left(\Omega
_0,\omega _4^{z_1},\omega _7^{z_1},\omega _0^z,\omega _0^y;s\right)+\frac{38}{3} J\left(\Omega _0,\omega _4^{z_1},\omega _7^{z_1},\omega _0^z,\omega
_0^z;s\right) \\ +\frac{38}{3} J\left(\Omega _0,\omega _4^{z_1},\omega _7^{z_1},\omega _y^z,\omega _0^y;s\right)-\frac{38}{3} J\left(\Omega _0,\omega _4^{z_1},\omega
_7^{z_1},\omega _y^z,\omega _0^z;s\right)+\frac{38}{3} J\left(\Omega _0,\omega _4^{z_1},\omega _7^{z_1},\omega _{\bar{y}}^z,\omega _0^y;s\right) \\ +\frac{38}{3}
J\left(\Omega _0,\omega _4^{z_1},\omega _7^{z_1},\omega _{\bar{y}}^z,\omega _0^z;s\right)+\frac{9}{2} J\left(\Omega _0,\omega _4^{z_1},\omega _9^{z_1},\omega
_0^y,\omega _0^y;s\right)+\frac{9}{2} J\left(\Omega _0,\omega _4^{z_1},\omega _9^{z_1},\omega _0^z,\omega _0^y;s\right) \\ -\frac{9}{2} J\left(\Omega _0,\omega
_4^{z_1},\omega _9^{z_1},\omega _0^z,\omega _0^z;s\right)-\frac{9}{2} J\left(\Omega _0,\omega _4^{z_1},\omega _9^{z_1},\omega _y^z,\omega
_0^y;s\right)+\frac{9}{2} J\left(\Omega _0,\omega _4^{z_1},\omega _9^{z_1},\omega _y^z,\omega _0^z;s\right) \\ -\frac{9}{2} J\left(\Omega _0,\omega _4^{z_1},\omega
_9^{z_1},\omega _{\bar{y}}^z,\omega _0^y;s\right)-\frac{9}{2} J\left(\Omega _0,\omega _4^{z_1},\omega _9^{z_1},\omega _{\bar{y}}^z,\omega _0^z;s\right)\Big] + \OO(\ep)\, .
\end{multline}

\begin{multline}
e^{2\ep\gamma_E}j_2^{\kite}(0,1,1,1,1)= e^{2\ep\gamma_E}j_3^{\kite}(0,1,1,1,1)
=\frac{1}{2 \varepsilon ^2}+\frac{z_1 J\left(\Omega _0,\omega
_4^{z_1};s\right)+\frac{5}{2}}{\varepsilon } \\ +\frac{1}{24} \left(-162 S_2+5 \pi ^2+132\right)-J\left(\Omega _0,\omega _0^z;s\right)+\frac{5}{96} \left(-12+\pi ^2\right) z_1 J\left(\Omega _0,\omega
_1^{z_1};s\right) \\ -\frac{2}{9} \left(-33+\pi ^2\right) z_1 J\left(\Omega _0,\omega _4^{z_1};s\right)+\frac{19}{72} \left(-24+\pi ^2\right) z_1 J\left(\Omega _0,\omega _7^{z_1};s\right)-\frac{3}{32} \left(-12+\pi
^2\right)z_1 J\left(\Omega _0,\omega _9^{z_1};s\right) \\ -\frac{5}{2} J\left(\Theta _1,\omega _0^y;s\right)+\frac{5}{2} J\left(\Theta _1,\omega
_0^z;s\right)+\frac{5}{2} J\left(\Theta _2,\omega _0^y;s\right)+\frac{5}{2} J\left(\Theta _2,\omega _0^z;s\right)-\frac{5}{2} J\left(\Theta _3,\omega
_0^z;s\right)+\frac{5}{2} J\left(\Theta _4,\omega _0^z;s\right) \\ -\frac{3}{2} J\left(\Theta _y,\omega _0^y;s\right)+\frac{3}{2} J\left(\Theta _y,\omega
_0^z;s\right)+\frac{3}{2} J\left(\Theta _{\bar{y}},\omega _0^y;s\right)  +\frac{3}{2} J\left(\Theta _{\bar{y}},\omega _0^z;s\right)+\frac{3}{2} J\left(\Omega
_0,\omega _0^y,\omega _0^y;s\right) \\ +\frac{3}{2} J\left(\Omega _0,\omega _0^z,\omega _0^y;s\right)-\frac{3}{2} J\left(\Omega _0,\omega _0^z,\omega
_0^z;s\right)+\frac{5}{8} z_1 J\left(\Omega _0,\omega _1^{z_1},\omega _0^z;s\right)-z_1 J\left(\Omega _0,\omega _4^s,\omega _4^{z_1};s\right)-2
z_1J\left(\Omega _0,\omega _4^{z_1},\omega _0^z;s\right) \\ -\frac{9}{8} z_1 J\left(\Omega _0,\omega _9^{z_1},\omega _0^z;s\right)-\frac{3}{2}
J\left(\Omega _0,\omega _y^z,\omega _0^y;s\right)+\frac{3}{2} J\left(\Omega _0,\omega _y^z,\omega _0^z;s\right)-\frac{3}{2} J\left(\Omega _0,\omega
_{\bar{y}}^z,\omega _0^y;s\right)-\frac{3}{2} J\left(\Omega _0,\omega _{\bar{y}}^z,\omega _0^z;s\right) \\ -\frac{2}{3} z_1 J\left(\Omega _0,\omega
_{4,-1}^{z_1},\omega _0^z;s\right)+\frac{2}{3} z_1 J\left(\Omega _0,\omega _{4,1}^{z_1},\omega _0^z;s\right)+\frac{19}{6} z_1 J\left(\Omega _0,\omega
_{7,-1}^{z_1},\omega _0^z;s\right)-\frac{19}{6} z_1 J\left(\Omega _0,\omega _{7,1}^{z_1},\omega _0^z;s\right) \\ -\frac{5}{8} z_1 J\left(\Omega _1,\omega
_{1,y}^{z_1},\omega _0^y;s\right)+\frac{5}{8} z_1 J\left(\Omega _1,\omega _{1,y}^{z_1},\omega _0^z;s\right)+\frac{8}{3} z_1 J\left(\Omega _1,\omega
_{4,y}^{z_1},\omega _0^y;s\right)-\frac{8}{3} z_1 J\left(\Omega _1,\omega _{4,y}^{z_1},\omega _0^z;s\right)\\ -\frac{19}{6} z_1 J\left(\Omega _1,\omega
_{7,y}^{z_1},\omega _0^y;s\right)+\frac{19}{6} z_1 J\left(\Omega _1,\omega _{7,y}^{z_1},\omega _0^z;s\right)+\frac{9}{8} z_1 J\left(\Omega _1,\omega
_{9,y}^{z_1},\omega _0^y;s\right)-\frac{9}{8} z_1 J\left(\Omega _1,\omega _{9,y}^{z_1},\omega _0^z;s\right)\\+\frac{5}{8} z_1 J\left(\Omega _2,\omega
_{1,\bar{y}}^{z_1},\omega _0^y;s\right)+\frac{5}{8} z_1 J\left(\Omega _2,\omega _{1,\bar{y}}^{z_1},\omega _0^z;s\right)-2 z_1 J\left(\Omega _2,\omega
_{4,\bar{y}}^{z_1},\omega _0^y;s\right)-2 z_1J\left(\Omega _2,\omega _{4,\bar{y}}^{z_1},\omega _0^z;s\right)\\ -\frac{9}{8} z_1 J\left(\Omega _2,\omega
_{9,\bar{y}}^{z_1},\omega _0^y;s\right)-\frac{9}{8}z_1 J\left(\Omega _2,\omega _{9,\bar{y}}^{z_1},\omega _0^z;s\right)+\frac{2}{3} z_1 J\left(\Omega
_3,\omega _{4,1}^{z_1},\omega _0^z;s\right)+\frac{1}{3} z_1 J\left(\Omega _3,\omega _{4,y}^{z_1},\omega _0^y;s\right)\\ -\frac{1}{3} z_1 J\left(\Omega
_3,\omega _{4,y}^{z_1},\omega _0^z;s\right)-\frac{1}{3} z_1 J\left(\Omega _3,\omega _{4,\bar{y}}^{z_1},\omega _0^y;s\right)-\frac{1}{3} z_1
J\left(\Omega _3,\omega _{4,\bar{y}}^{z_1},\omega _0^z;s\right)-\frac{19}{6} z_1 J\left(\Omega _3,\omega _{7,1}^{z_1},\omega _0^z;s\right)\\ -\frac{19}{12}
z_1 J\left(\Omega _3,\omega _{7,y}^{z_1},\omega _0^y;s\right)+\frac{19}{12} z_1 J\left(\Omega _3,\omega _{7,y}^{z_1},\omega
_0^z;s\right)+\frac{19}{12} z_1 J\left(\Omega _3,\omega _{7,\bar{y}}^{z_1},\omega _0^y;s\right)+\frac{19}{12} z_1 J\left(\Omega _3,\omega
_{7,\bar{y}}^{z_1},\omega _0^z;s\right)\\ +\frac{2}{3} z_1 J\left(\Omega _4,\omega _{4,-1}^{z_1},\omega _0^z;s\right)+\frac{1}{3} z_1 J\left(\Omega
_4,\omega _{4,y}^{z_1},\omega _0^y;s\right)-\frac{1}{3} z_1 J\left(\Omega _4,\omega _{4,y}^{z_1},\omega _0^z;s\right) -\frac{1}{3} z_1 J\left(\Omega
_4,\omega _{4,\bar{y}}^{z_1},\omega _0^y;s\right)\\-\frac{1}{3} z_1 J\left(\Omega _4,\omega _{4,\bar{y}}^{z_1},\omega _0^z;s\right)-\frac{19}{6} z_1
J\left(\Omega _4,\omega _{7,-1}^{z_1},\omega _0^z;s\right)-\frac{19}{12} z_1 J\left(\Omega _4,\omega _{7,y}^{z_1},\omega _0^y;s\right)+\frac{19}{12}
z_1 J\left(\Omega _4,\omega _{7,y}^{z_1},\omega _0^z;s\right)\\ +\frac{19}{12} z_1 J\left(\Omega _4,\omega _{7,\bar{y}}^{z_1},\omega
_0^y;s\right)+\frac{19}{12} z_1 J\left(\Omega _4,\omega _{7,\bar{y}}^{z_1},\omega _0^z;s\right)+\frac{5}{8} z_1 J\left(\Omega _0,\omega
_1^{z_1},\omega _0^y,\omega _0^y;s\right)+\frac{5}{8} z_1 J\left(\Omega _0,\omega _1^{z_1},\omega _0^z,\omega _0^y;s\right)\\-\frac{5}{8} z_1
J\left(\Omega _0,\omega _1^{z_1},\omega _0^z,\omega _0^z;s\right) -\frac{5}{8} z_1 J\left(\Omega _0,\omega _1^{z_1},\omega _y^z,\omega
_0^y;s\right)+\frac{5}{8} z_1 J\left(\Omega _0,\omega _1^{z_1},\omega _y^z,\omega _0^z;s\right) -\frac{5}{8} z_1 J\left(\Omega _0,\omega
_1^{z_1},\omega _{\bar{y}}^z,\omega _0^y;s\right)\\-\frac{5}{8}z_1 J\left(\Omega _0,\omega _1^{z_1},\omega _{\bar{y}}^z,\omega _0^z;s\right)-\frac{8}{3}
z_1 J\left(\Omega _0,\omega _4^{z_1},\omega _0^y,\omega _0^y;s\right)-\frac{8}{3} z_1 J\left(\Omega _0,\omega _4^{z_1},\omega _0^z,\omega
_0^y;s\right)+\frac{8}{3}z_1 J\left(\Omega _0,\omega _4^{z_1},\omega _0^z,\omega _0^z;s\right)\\ +\frac{8}{3}z_1 J\left(\Omega _0,\omega
_4^{z_1},\omega _y^z,\omega _0^y;s\right)-\frac{8}{3}z_1 J\left(\Omega _0,\omega _4^{z_1},\omega _y^z,\omega _0^z;s\right)+\frac{8}{3}z_1
J\left(\Omega _0,\omega _4^{z_1},\omega _{\bar{y}}^z,\omega _0^y;s\right)+\frac{8}{3}z_1 J\left(\Omega _0,\omega _4^{z_1},\omega _{\bar{y}}^z,\omega
_0^z;s\right)\\+\frac{19}{6} z_1 J\left(\Omega _0,\omega _7^{z_1},\omega _0^y,\omega _0^y;s\right)+\frac{19}{6} z_1 J\left(\Omega _0,\omega
_7^{z_1},\omega _0^z,\omega _0^y;s\right)-\frac{19}{6}z_1 J\left(\Omega _0,\omega _7^{z_1},\omega _0^z,\omega _0^z;s\right)-\frac{19}{6}z_1
J\left(\Omega _0,\omega _7^{z_1},\omega _y^z,\omega _0^y;s\right)\\ +\frac{19}{6} z_1 J\left(\Omega _0,\omega _7^{z_1},\omega _y^z,\omega
_0^z;s\right)-\frac{19}{6} z_1 J\left(\Omega _0,\omega _7^{z_1},\omega _{\bar{y}}^z,\omega _0^y;s\right)-\frac{19}{6} z_1 J\left(\Omega _0,\omega
_7^{z_1},\omega _{\bar{y}}^z,\omega _0^z;s\right)-\frac{9}{8} z_1 J\left(\Omega _0,\omega _9^{z_1},\omega _0^y,\omega _0^y;s\right)\\-\frac{9}{8} z_1
J\left(\Omega _0,\omega _9^{z_1},\omega _0^z,\omega _0^y;s\right)+\frac{9}{8} z_1 J\left(\Omega _0,\omega _9^{z_1},\omega _0^z,\omega
_0^z;s\right)+\frac{9}{8} z_1 J\left(\Omega _0,\omega _9^{z_1},\omega _y^z,\omega _0^y;s\right)-\frac{9}{8} z_1 J\left(\Omega _0,\omega
_9^{z_1},\omega _y^z,\omega _0^z;s\right)\\+\frac{9}{8}z_1 J\left(\Omega _0,\omega _9^{z_1},\omega _{\bar{y}}^z,\omega _0^y;s\right)+\frac{9}{8} z_1
J\left(\Omega _0,\omega _9^{z_1},\omega _{\bar{y}}^z,\omega _0^z;s\right) + \OO (\ep), .
\end{multline}

\bibliographystyle{hieeetr}
\bibliography{litr}

\end{document}